\documentclass[12pt,a4paper]{article}
\pdfoutput=1
\usepackage[a4paper]{geometry}
\usepackage{amsmath}
\usepackage{amsfonts}
\usepackage{amssymb}
\usepackage{dsfont}
\usepackage{cite}
\usepackage{hyperref}
\usepackage{multirow}
\usepackage{xcolor}
\usepackage{soul}
\usepackage[font=small, font+=it, format=hang, margin={1cm, 1cm}]{caption}

\usepackage{tikz}
\usetikzlibrary{calc,decorations.markings}
\usetikzlibrary{snakes}
\usepackage{rotating}
\usepackage{comment}
\usepackage{empheq}
\usepackage{braket}
\usepackage{tensor}
\usepackage{url}

\usepackage{cancel} 
\usepackage{empheq}
\usepackage{mathrsfs}
\usepackage{csquotes}
\usepackage{multicol}
\usepackage{stmaryrd}

\newcommand{\pd}{\text{d}}
\newcommand{\p}{\partial}
\newcommand{\sgn}{\text{sgn}}
\newcommand{\hbc}{\hat{\boldsymbol{c}}}
\newcommand{\bx}{\boldsymbol{x}}
\newcommand{\by}{\boldsymbol{y}}

\newcommand{\R}{\mathbb{R}}

\newcommand{\sM}{\mathcal{M}}
\newcommand{\diag}{\text{diag}}
\newcommand{\sO}{\mathcal{O}}
\newcommand{\bphi}{\boldsymbol{\Phi}}
\newcommand{\bpsi}{\boldsymbol{\Psi}}
\newcommand{\bzeta}{\boldsymbol{\zeta}}
\newcommand{\bchi}{\boldsymbol{\chi}}
\newcommand{\cA}{\mathcal{A}}
\newcommand{\scri}{\mathcal{I}}

\newcommand{\const}{\text{const}}

\newcommand{\bv}{\boldsymbol{v}}
\newcommand{\bR}{\boldsymbol{R}}

\newcommand{\fa}{\mathsf{a}}
\newcommand{\fb}{\mathsf{b}}

\newcommand{\fc}{\mathsf{c}}
\newcommand{\tr}{\text{tr}}
\newcommand{\bth}{\boldsymbol{\theta}}

\allowdisplaybreaks[2]
\numberwithin{equation}{section}

\definecolor{darkgreen}{rgb}{0.0, 0.55, 0.1}

\begin{document}

\begin{titlepage}

    \begin{flushright}\vspace{2cm}
        {\small
  LMU-ASC 25/22\\
  MS-TP-22-22
  }
    \end{flushright}
    \bigskip
    \vspace{2cm}

    \begin{center}
        %
        {\Huge Asymptotic symmetries and memories \\ of gauge theories in FLRW spacetimes\vspace{15mm}}
        
        \bigskip
        
            Mart\'in Enr\'iquez Rojo$^{\dagger}$\footnote{martin.enriquez@physik.lmu.de}
            and
            Tobias Schr\"oder$^{\dagger *}$\footnote{schroeder.tobias@uni-muenster.de}
            \vspace{20pt}
        \bigskip
        \medskip
        
        \textit{{}$^\dagger$ Arnold Sommerfeld Center for Theoretical Physics\\
Ludwig-Maximilians-Universit\"at\\
Theresienstraße 37\\
80333 M\"unchen, Germany}\\
\vspace{5mm}
\textit{{}$^*$ Institute of Theoretical Physics\\
University of M\"unster\\
Wilhelm-Klemm-Straße 9\\
48149 M\"unster, Germany}

        \vspace{20mm}
    

\begin{abstract}
\vspace{1em} 
In this paper, we investigate the asymptotic structure of gauge theories in decelera-ting and spatially flat Friedmann–Lema\^itre–Robertson–Walker universes. Firstly, we tho-roughly explore the asymptotic symmetries of electrodynamics in this background, which reveals a major inconsistency already present in the flat case. Taking advantage of this treatment, we derive the associated memory effects, discussing their regime of validity and differences with respect to their flat counterparts. Next, we extend our analysis to non-Abelian Yang-Mills, coupling it dynamically and simultaneously to a Dirac spinor and a complex scalar field. Within this novel setting, we examine the possibility of constructing Poisson superbrackets based on the covariant phase space formalism.
\end{abstract}
        
    \end{center}
    \setcounter{footnote}{0}
\end{titlepage}

\tableofcontents

\section{Introduction}  
\label{Sec:Intro}

The pioneer work of Bondi, van der Burg, Metzner and Sachs (BMS) in the 1960s opened the door to the investigation of asymptotic symmetries in asymptotically flat spacetimes \cite{BMS1, BMS2, BMS3}. The latter are characterized by placing asymptotic fall-off conditions on the gravitational radiation at null infinity, thereby restricting the allowed geometries. The question addressed then was which diffeomorphisms leave these boundary conditions untouched. Quite naturally, the expectation was that the symmetry group of general relativity would reduce to the isometry group of flat Minkowski spacetime. Surprisingly, they found instead the infinite-dimensional BMS group which includes supertranslations on top of the Poincaré group.

In the recent years, several extensions of the BMS group have been introduced \cite{Barnich:2009se,Barnich:2010eb,CampigliaLaddha,Campiglia:2015yka,WeylBMS}, and a fascinating connection between the asymptotic symmetries, Weinberg's soft graviton theorems \cite{WeinbergOld, Weinbergpaper} and gravitational memory effects \cite{ZeldovichPolnarev,Christodoulou} has been unveiled under the name of the gravitational infrared triangle \cite{Stromingerseminal1,Stromingerseminal2,Strominger}. On a technical level, the connection between soft theorems and memory effects is nothing else than a Fourier transform from momentum to position space. Memory effects can also be expressed in terms of a difference between inequivalent radiative vacua, namely before and after the passage of the gravitational waves producing the memories. BMS symmetries relate such inequivalent vacua. Thus, memory effects are a priori physically observable consequences of the more abstract asymptotic symmetries.

Such an infrared structure is not exclusive of the gravitational interaction. In fact, analogous soft theorems for quantum electrodynamics date back to a 1937 article by Bloch and Nordsieck \cite{NordsieckBloch}, while the work of \cite{BieriChenYau,Bieri} found the corresponding memory effect in pure electrodynamics. What remains to be discussed are the asymptotic symmetries. By analogy with gravity, in gauge theories, one can restrict the allowed gauge field configurations to certain asymptotic decay behaviors. One would then expect the asymptotic symmetries to be the subset of the gauge symmetries (in the case of electrodynamics, the $U(1)$ symmetry) which do not violate the imposed boundary conditions. In fact, this analogy naturally works out. The systematic investigation of these symmetries, comprising those gauge transformations that do not vanish at null infinity, was initiated with Strominger's seminal article \cite{StromingerYangMills} in 2014.

As a consequence, the infrared triangle of gauge theories and, in particular, electrodynamics, resembles the one of gravity. The relation between asymptotic gauge symmetries and soft theorems was already introduced in \cite{StromingerYangMills}. The link to memory effects was, in the case of electrodynamics, already apparent from Bieri's and Garfinkle's article \cite{Bieri} as soon as asymptotic gauge symmetries were explored. The connection was then clarified in \cite{Pasterski} which also demonstrated the interconnection between soft photon theorems and memory effects. Similarly to the two other corners of the infrared triangle, memory effects are not only a feature of electrodynamics but also show up in non-Abelian gauge theories \cite{Pate}.

So far, we referred either to asymptotically flat spacetimes in the case of gravity or to a Minkowski background in the case of gauge theories. Besides, we recall that the asymptotic analysis considers observers located at extremely large distances from all sources. While such studies are quite interesting on their own, observations suggest that the universe we live in is, at very large scales, best described by a spatially flat expanding FLRW spacetime, currently undergoing accelerated expansion \cite{Bennett, Supernova}.
It is surprising that in spite of this, fairly few investigations on the infrared structure have been carried out in the context of FLRW spacetimes \cite{Hawking:1968qt,Shiromizu:1999iq,Ferreira:2016hee,Compere:2019bua,Compere:2020lrt,Chrusciel:2021ttc,Hinterbichler:2013dpa,Mirbabayi:2016xvc,Pajer:2017hmb,Hamada:2018vrw,Bieri:2015jwa,Tolish:2016ggo,Chu:2016qxp,Chu:2016ngc,Bieri:2017vni,Hamada:2017gdg,Chu:2021apx,Jokela:2022rhk,Bonga,Heckelbacher,Rojo, Olivieri}. On the technical side, most of the aforementioned analyses in Minkowski space  were dealing with null infinity due to the latter being the asymptotic region reached by radiation. Therefore, as the first step towards this phenomenologically more relevant case, it would be advantageous to pick out those FLRW spacetimes which possess a null infinity. The only FLRW spacetimes (with linear equation of state) comprising this feature are the decelerating and spatially flat ones \cite{Harada1, Harada}. In this class of spacetimes, the cosmologically relevant cases of matter and radiation dominated universes are contained.

The analysis of BMS has been recently extended to these spacetimes by \cite{Bonga} and \cite{Heckelbacher}, following independent approaches. Later, the discussion has been extensively expanded in \cite{Rojo, Olivieri} and further complemented by \cite{Enriquez-Rojo:2021rtv,EnriquezRojo:2022swn} \footnote{The interested reader can find a compact and detailed treatment of asymptotic symmetries in FLRW universes for the gravitational interaction in the dissertation of one of the authors \cite{EnriquezRojo:2022swn}.}. 
While these studies focus on the asymptotic structure of gravity, to the best of our knowledge, there are no such explorations for the case of gauge theories in cosmological spacetimes. 

The central motivation of the present work is two-fold. On the one hand, we aim to fill this gap by exploring asymptotic symmetries and memory effects of electrodynamics and Yang-Mills theory in decelerating and spatially flat FLRW cosmologies. On the other hand, we use this opportunity to complement the existing literature with an exhaustive and computationally explicit treatment of the topic and the addition of a novel and challenging setting coupling Yang-Mills to both bosons and fermions, which introduces fermions in the phase space \footnote{To our knowledge, such an analysis was only partially considered for the case of flat spacetime, in particular without fermions \cite{HeMitraCov}. In \cite{MitraPhD}, the asymptotic gauge symmetries of $\mathcal{N}=1$ supersymmetric QED have been explored in flat background without attempting to define a consistent phase space.}.

Let us briefly summarize the results obtained within this manuscript:
\begin{itemize}
    \item We introduce a directed volume form on null infinity via a simple regularization procedure for both Minkowski and FLRW spacetimes in section \ref{Sec:decFLRW}. Although the flat spacetime result was used in previous works, we contribute to the existing literature with an explicit derivation.
    \item We investigate the radial fall-offs of conformally coupled complex scalar and Dirac fields in FLRW backgrounds in section \ref{Subsec:currents}.
    \item We analyze in great detail the asymptotic symmetry corner of the FLRW infrared triangle for gauge theories in sections \ref{Sec:ED} and \ref{Sec:YM}. Especially, we point out a serious inconsistency concerning the introduction of an ``extended phase space" at null infinity and the definition of the soft photon field required to connect asymptotic symmetries with soft theorems and memory effects in sections \ref{Subsec:asympsym} and \ref{Subsec:YangMillsSymmetries}.
    \item We discuss gauge memory effects in FLRW spacetimes within section \ref{Subsec:Memories}, including a derivation of Liénard-Wiechert-like solutions for spatially flat FLRW spacetimes (appendix \ref{LienardWiechertLikeSolutionsforFLRW}).
    \item We explore the asymptotic behavior and phase space of a non-Abelian gauge field with dynamically coupled matter fields, i.e. conformally coupled complex scalar fields and Dirac fields, for a FLRW background in section \ref{subsec:fermionicps} and with the help of appendix \ref{AppendixExpansion}. In particular, we also attempt to construct the asymptotic phase space of the theory, observing an intriguing tension (when dynamical fermions are included) between the use of Poisson superbrackets and the phase space formulation.
\end{itemize}

This manuscript is organized as follows: In sections~\ref{Sec:decFLRW} and \ref{Sec:CPS}, we briefly describe decelerating and spatially flat FLRW spacetimes and review the covariant phase space formalism, respectively. In section~\ref{Sec:ED}, we explore the asymptotic symmetries of electrodynamics in these cosmological backgrounds and their associated memory effects. Our section~\ref{Sec:YM} is devoted to the study of asymptotic symmetries for Yang-Mills coupled to a complex scalar and a Dirac field in FLRW. Section~\ref{Sec:Conclusions} contains 
our summary and conclusions. The reader can find the conventions we use when dealing with spinors in appendix~\ref{Appendix:Spinorconventions}, as well as details on the asymptotic expansion of Yang-Mills coupled to dynamical matter in appendix~\ref{AppendixExpansion} and a thorough derivation of Liénard-Wiechert-like solutions for FLRW, necessary for our investigation of memory effects, in appendix~\ref{LienardWiechertLikeSolutionsforFLRW}.

\vspace{1em}

We furthermore mention that this paper is based on the master thesis \cite{tobias:2022}
of one of the authors, where further details and discussions can be found. 

\paragraph{Notation:}  We denote Minkowski spacetime by $\mathbb{M}$ and the corresponding future null infinity by $\scri^+$. We signify the considered FLRW spacetimes by $\mathcal{M}$ and their future null infinity by $\scri$. The components of a vector field $V$ are defined via $V=V^u \frac{\p}{\p u}+V^r \frac{\p}{\p r}+ V^A \frac{\p }{\p \theta^A}$. Angular dependencies are either denoted by $\bth$ or, alternatively, by specifying a corresponding unit $3$-vector on the sphere: $\hat{\bx}$, such that $\hat{x}^A=\theta^A$. The round metric on the sphere is $\gamma_{AB}$. For the covariant derivative on $S^2$ we use $D_A$ and for the volume form $\pd \Gamma$. The Dirac delta on the sphere is denoted as $\delta_{S^2}(\bth, \bth')$ and normalized such that $\oint_{S^2} \delta_{S^2}(\bth, \bth')\pd \Gamma =1$. Exterior derivatives and exterior products on the spacetime manifold are represented by $\pd$ and $\wedge$, while they are denoted on the phase space (configuration/solution space) by $\delta$ and $\curlywedge$. For spinors, we use notations and conventions introduced in section \ref{Subsubsec:Dirac} and appendix \ref{Appendix:Spinorconventions}. When working with $SU(\mathcal{N})$-Yang-Mills theory, upper indices $\fa, \fb\in \Set{1, \dots, \mathcal{N}^2-1}$ of the gauge fields are associated with the adjoint representation, whereas lower indices $\mathfrak{i},\mathfrak{j}\in \Set{1,\dots,\mathcal{N}}$ of the matter fields are associated with the fundamental representation.


\section{Decelerating and spatially flat FLRW spacetimes}  
\label{Sec:decFLRW}

According to the cosmological principle, our universe is spatially homogeneous and isotropic on large scales. At each point of an FLRW spacetime, there exists an observer who sees a spatially homogeneous and isotropic universe. Such an observer is called comoving. Correspondingly, the metric of spatially flat FLRW spacetimes sourced by a fluid with an equation of state parameter $w=p/\rho$ is given, in comoving coordinates, by
\begin{align}
\pd s^2=\pd t^2-a(t)^2\pd \bx^2 \ , \ \ \ \ a(t)=\left(\frac{t}{t_0}\right)^{\frac{2}{3(w+1)}} \ ,
\label{FLRWStandard}
\end{align}
with $t\in (0, \infty)$ and $\bx^i\in \R$.

These metrics are related to the Minkowski metric by a Weyl transformation. In fact, using the conformal time $\pd \eta= \frac{\pd t}{a(t)}$ and introducing retarded Bondi coordinates $(u,r,\bth)$, with $u=\eta-\sqrt{x^ix_i}$ and $r=\sqrt{x^ix_i}$, the spatially flat FLRW metric reads
\begin{empheq}{align}
\pd s^2=\left(\frac{u+r}{\eta_0}\right)^{2s}\left(\pd u^2+2\pd u \pd r -r^2\gamma_{AB}\pd\theta^A \pd \theta^B\right) \ , 
\label{Metric}
\end{empheq}
where $\eta_0=\eta(t_0)$ is the conformal time at present, $\gamma_{AB}$ is the round metric on the sphere and we introduced the parameter $s=\frac{2}{1+3w}$. 

These spacetimes can be divided into decelerating ($s>0$) and accelerating ($s<0$). We will restrict our considerations to the former case, the reason being that we are interested in asymptotic symmetries at null infinity which exists only for the decelerating case (cf. \cite{Mukhanov,Harada1, Harada}). Among them, the physically relevant matter ($s=2$) and radiation ($s=1$) dominated universes are included. Besides, the metric clearly presents a Big Bang singularity at $\eta\to 0$.
    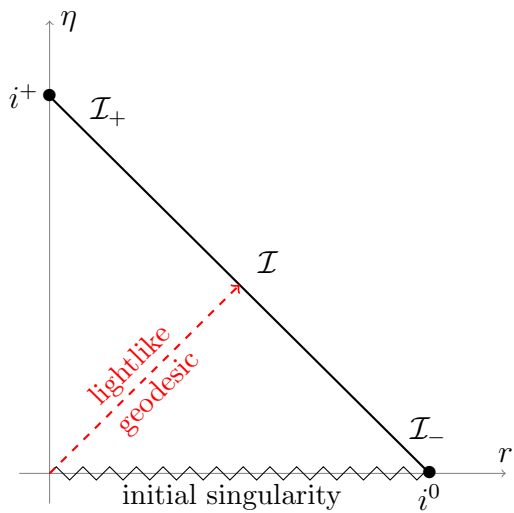
\begin{figure}[h]
		\begin{center}
				\begin{tikzpicture}[scale=2.0]
					\draw[help lines,->] (-0.2,0) -- (3,0) coordinate (raxis);
					\draw[help lines,->] (0,-0.2) -- (0,3) coordinate (taxis);
					
					\draw[thick] (2.5,0) -- (0,2.5);
					\draw[snake] (0,0) -- (2.5,0);
					\draw[thick, dashed, red, ->] (0,0) -- (1.25,1.25);
					
					\node[above] at (raxis) {$r$};
					\node[right] at (taxis) {$\eta$};
					\node[below] at (1.2,0) {{\small initial singularity}};
					\node[align=center] at (0.625,0.625) [rotate=45] {{\small\textcolor{red}{lightlike}}\\{\small\textcolor{red}{geodesic}}};
					\node at (0,2.5) {$\bullet$};
					\node[left] at (0,2.5) {$i^+$};
					\node[right] at (0.2,2.4) {$\mathcal{I}_+$};
					\node at (2.5,0) {$\bullet$};
					\node[below] at (2.5,0) {$i^0$};
					\node[right] at (2.3,0.28) {$\mathcal{I}_-$};
					\node[right] at (1.3,1.4) {$\mathcal{I}$};
				\end{tikzpicture}
		\end{center}
		\caption{Conformal diagram of decelerating and spatially flat FLRW spacetimes. Points on $\scri$ are accessible by keeping $u$ constant and taking the limit $r\to \infty$. Adopting next the limits $u\to \pm \infty$, one reaches the early and late time boundaries $\mathcal{I}_{\pm}$.}
		\label{fig:conformal_diagram}
	\end{figure}

Later on, it will be necessary to integrate vector fields over $\mathcal{I}$. As a consequence, we will now compute the directed volume form on future null infinity.

Before we continue, we would like to emphasize that the approach we will follow here is similar to that of \cite{Wilczek}, where a null hypersurface (black hole horizon) is approached by considering a timelike hypersurface (stretched horizon) on which the computation is performed, such that only in the end one takes the limit to the actual null hypersurface. The flat limit of the result we will derive is compatible with that partially used in the flat literature (see e.g. \cite{MitraPhD}). Nevertheless, a detailed derivation is missing, to the best of our knowledge, and we aim to fill this gap.

We start by constructing the normal vector to $\scri$. For any hypersurface $\Sigma\subset M$, the normal vector $n^\mu$ can be found by expressing $\Sigma$ as the zero set of a function $f:M\to \R$. If $\p_\mu f\neq 0$, the normal covector can be found as $n_\mu=\p_\mu f$. $\scri$ is characterized by $v\vert_{\scri}\equiv (\eta+r)\vert_{\scri}=(u+2r)\vert_{\scri}=\const$, even though $v\vert_{\scri}$ is infinite. Thereby, one finds $\scri$ as the zero set of the function $f(u,r)=u+2r-v\vert_{\scri}$. Hereof, we find $n_\mu=\p_\mu f=\delta_\mu^u + 2\delta_\mu^r\neq 0$ and, accordingly, $n^\mu=2\delta_u^\mu-\delta^\mu_r$ which is also future directed. However, this is a null vector and, therefore, we cannot normalize it, fixing an arbitrary prefactor. We will soon come back to this issue.

Next, we want to find the directed surface element $\pd \Sigma^\mu$. For any non-null hypersurface, this can be done as follows: Let $M$ be a smooth manifold with volume form $\varpi_M$ and hypersurface $\Sigma\subset M$. We denote the inclusion as $\iota_\Sigma: \Sigma\to M$. Let $N$ be the unit normal vector field along $\Sigma$. Then, the induced volume form is given by the pullback \cite{Lee} 
\begin{align}
\pd \Sigma=\iota_\Sigma^* (i_N\varpi_M),\label{inducedvolumeformgeneral}
\end{align}
where $i_N$ denotes the interior product. The directed volume form can then be found as $\pd \Sigma^\mu=N^\mu \pd \Sigma$. Unfortunately, for null surfaces the volume form obtained in this way vanishes and, as previously mentioned, the normal vector is not normalizable. 
To proceed, we will use a regularization procedure. We start with timelike \footnote{In principle, we could have also started with spacelike hypersurfaces. The results deviate only by an overall minus sign which we could also insert by hand.} hypersurfaces, then use the above procedure to find the directed volume form and take the limit to the null surface only in the end.
As mentioned before, $\scri$ can be found as the zero set of the function $f(u,r)=u+2r-v\vert_\scri$ where $v\vert_\scri=\infty=\const.$ We define timelike surfaces as the zero sets of the functions
\begin{align}
F(u,r,\beta)=(2-\beta)u+2r-v\vert_\scri && \text{and} && \beta>1 ,
\end{align}
thus, approaching $\scri$ in the limit $\beta\to 1^+$. The corresponding normal unit-covector is $N_\mu(\beta)=\frac{1}{a(u,r)\sqrt{\beta-1}}\left(\left(1-\frac{\beta}{2}\right)\delta^u_\mu +\delta^r_\mu\right)$ and the normal vector is $N^\mu(\beta)=\frac{1}{a(u,r)\sqrt{\beta-1}}\left(\delta^\mu_u-\frac{\beta}{2}\delta^\mu_r\right)$. With the volume form $\varpi_\mathcal{M}=a^4(u,r)r^2 \pd u \wedge \pd r \wedge \pd \Gamma$, we find via \eqref{inducedvolumeformgeneral}
\begin{align}
\pd \Sigma(\beta)=\lim_{r\to\infty} \frac{a^3(u,r)r^2 (\beta-1)}{\sqrt{\beta-1}} \pd u\wedge \pd \Gamma. 
\end{align}
Multiplying by $N^\mu(\beta)$, we get $\pd \Sigma^\mu(\beta)= \lim_{r\to \infty} a^2(u,r)r^2 \left(\delta^\mu_u-\frac{\beta}{2}\delta^\mu_r\right)\pd u\wedge \pd \Gamma$ and are now able to take the limit $\beta\to 1^+$ to obtain the directed volume form on $\scri$
\begin{empheq}{align}
\pd \Sigma^\mu=\lim_{r\to \infty} a^2(u,r) r^2 \left(\delta^\mu_u-\frac{1}{2}\delta^\mu_r\right)\pd u\wedge \pd \Gamma.
\end{empheq}

\section{Review of covariant phase space formalism}  
\label{Sec:CPS}

In this section, we present a short revision of the covariant phase space formalism \cite{LeeWald, IyerWald, WaldZoupas,CrnkovicWitten, Crnkovic} which will be needed in the following sections. In the course of this discussion, we mainly follow \cite{HarlowWu,HeMitraCov,Gieres} and recommend them to the interested reader for further details.

A phase space is a differentiable manifold $\Gamma$ endowed with a symplectic form $\Omega$, the latter being a 2-form satisfying
\begin{itemize}
    \item[(i)] $\delta \Omega=0$, 
    \item[(ii)] $\forall X\in T\Gamma:$ If $i_X\Omega=0$ then $X=0$. 
\end{itemize}
We denote the exterior derivative on $\Gamma$ by $\delta$ in order to distinguish it from the exterior derivative $\pd$ on spacetime. $T\Gamma$ denotes the tangent bundle over $\Gamma$ and in the following $T^*\Gamma$ is its dual bundle. \\
We can now define a function $\hat{\Omega}: T\Gamma \to T^*\Gamma, \; X\mapsto -i_X \Omega$ which is injective, since $\Omega$ is bilinear and non-degenerate. For a finite-dimensional phase space, we can, therefore, invert $\hat{\Omega}$ where the inverse is the map $\hat{\Omega}^{-1}: T^*\Gamma\to T\Gamma$ defined by
\begin{align}
\forall X\in T\Gamma: \hat{\Omega}^{-1}(-i_X\Omega)=X && \text{and} &&\forall \chi \in T^*\Gamma: i_{\hat{\Omega}^{-1}(\chi)}\Omega=-\chi.
\end{align}
This allows us to define the antisymmetric bilinear map 
\begin{align}
\Omega^{-1}: T^*\Gamma\times T^*\Gamma\to \mathfrak{F}(\Gamma), \; (\chi,\psi)\mapsto \Omega^{-1}(\chi, \psi)=\chi(\hat{\Omega}^{-1}(\psi)),
\end{align}
where $\mathfrak{F}(\Gamma)$ is the space of smooth functions on $\Gamma$. 
Thus, one can easily check that it satisfies
\begin{align}
\forall X,Y\in T\Gamma: \Omega^{-1}(i_X \Omega, i_Y\Omega)=-\Omega(X,Y).
\end{align}
This enables us to define for any function $\mathfrak{g}\in \mathfrak{F}(\Gamma)$  a vector field $X_\mathfrak{g}$ (called Hamiltonian vector field) by its action on functions as
\begin{align}
\forall \mathfrak{f}\in \mathfrak{F}(\Gamma): X_\mathfrak{g}(\mathfrak{f})=\Omega^{-1}(\delta \mathfrak{f}, \delta \mathfrak{g})
\end{align}
which one finds, due to the above definitions, to be equivalent to
\begin{align}
i_{X_\mathfrak{g}}\Omega =-\delta \mathfrak{g}. \label{phasespacevectorfielddef}
\end{align}
Finally, we define the Poisson bracket between any two functions $\mathfrak{f},\mathfrak{g}\in \mathfrak{F}(\Gamma)$ as
\begin{align}
\Set{\mathfrak{f},\mathfrak{g}}=-\Omega(X_\mathfrak{f}, X_\mathfrak{g}). \label{definitionPoissonbracket}
\end{align}
It is hence clear that the requirement of non-degeneracy in the definition of the symplectic form is needed to define Poisson brackets. Otherwise, $\Omega^{-1}$ does not exist and we cannot define $X_\mathfrak{g}$ for any function $\mathfrak{g}$. This is generally true. In our treatment, we are dealing on top of this with field theories and, correspondingly, infinite-dimensional phase spaces. In this case, the injectivity of $\hat{\Omega}$ does not automatically imply its surjectivity. Therefore, non-degeneracy is necessary but not sufficient to obtain Poisson brackets from $\Omega$. This problem can be quite subtle and we will address it once encountered. In order to describe the formalism, we will assume in the following that we have fixed this issue and $\hat{\Omega}^{-1}$ exists.

We will now review how the phase space is constructed for a given Lagrangian, which is the actual covariant phase space formalism. We consider a field theory on a spacetime manifold $M$ describing the dynamics of some fields $\phi^i$. All kinematically possible field configurations which satisfy the boundary conditions of the theory 
form the configuration space $\mathcal{C}$. The dynamics of the theory is described in terms of a Lagrangian form $L$ which depends on the fields and may also depend on an arbitrary (but finite) number of derivatives of the field, as well as on some background structure. The Lagrangian form is just the usual Lagrangian density times the volume form on $M$. It is, therefore, a $4$-form on $M$ and a scalar on $\mathcal{C}$. The variation (exterior derivative) of the Lagrangian can always be brought into the form \cite{LeeWald}
\begin{align}
\delta L =\sum_i E_i \delta \phi^i+\pd \Theta.\label{variationLagrangian}
\end{align}
The symplectic potential, $\Theta$, has to be a $3$-form on $M$ while it is a $1$-form on $\mathcal{C}$. It is only defined up to $\Theta \longrightarrow \Theta +\pd Y$ where $Y$ is a local $2$-form on $M$ and a $1$-form on $\mathcal{C}$. The dynamics of the theory is obtained by imposing the variation of the action to vanish up to boundary terms. Since $\delta S=\int_M \delta L$, this least action principle implies
\begin{align}
E_i(\set{\bar{\phi}^j})=0
\end{align}
which are the equations of motion. The subspace of field configurations $\bar{\phi}^i$ satisfying the equations of motion is called solution space $\mathcal{S}$.
Now, recall that $\Gamma$ is usually interpreted as the space of all possible different initial conditions on a Cauchy slice of constant time, which is a non-covariant notion. However, for a well-defined initial value problem, there is a bijection between $\Gamma$ and $\mathcal{S}$, so $\mathcal{S}$ can be used as the phase space, which would then be defined in a covariant manner. In order to really obtain a phase space, however, we still have to find the symplectic form for $\mathcal{S}$, which is approached as follows. From the symplectic potential $\Theta$ (now restricted to $\mathcal{S}$) we define a $2$-form on $\mathcal{S}$\cite{LeeWald}
\begin{align}
\omega=\delta \Theta,
\end{align}
called the symplectic current. Due to the ambiguity in $\Theta$, the symplectic current is only defined up to $\omega \longrightarrow \omega+\delta \pd Y$. Since $\omega$ is a $3$-form on $M$, we can naturally integrate it over a $3$-hypersurface in $M$. Thereby, we define the presymplectic form as the integral of $\omega$ over any Cauchy slice $\Sigma\subset M$ \cite{Lee}
\begin{align}
\tilde{\Omega}_\Sigma=\int_\Sigma \omega,  \label{omegadefuniquenon}
\end{align}
leaving an ambiguity $\tilde{\Omega}_\Sigma\longrightarrow \tilde{\Omega}_\Sigma + \int_{\p\Sigma} \delta Y$. In addition to the property $\delta \tilde{\Omega}_\Sigma=0$, satisfied by construction, a symplectic form must be non-degenerate which is not generally true for $\tilde{\Omega}_\Sigma$. As noticed earlier, the initial value problem must be well-defined in order to identify $\mathcal{S}$ and $\Gamma$. 
In gauge theories, this is not the case due to the gauge redundancy. This issue can be addressed by gauge fixing which maps equivalence classes of gauge fields related by a gauge transformation to their one representative satisfying the gauge condition. The restriction of $\tilde{\Omega}_\Sigma$ to $\Gamma$ is then the symplectic form, such that we have constructed the phase space. As mentioned before, while gauge fixing may be enough to make the presymplectic form non-degenerate, it might still be non-invertible on infinite dimensional phase spaces. We will encounter this issue in section \ref{Subsec:asympsym}.

Finally, let us assume that the theory exhibits a continuous symmetry, taking, for an infinitesimal parameter $\epsilon$, the form $\mathfrak{f}\longrightarrow \mathfrak{f}+\delta_\epsilon \mathfrak{f}$, where $\mathfrak{f}\in \mathfrak{F}(\Gamma)$. The charge $\mathcal{Q}[\epsilon]$ generating this symmetry is supposed to satisfy $\Set{\mathfrak{f}, \mathcal{Q}[\epsilon]}=\delta_\epsilon \mathfrak{f}$. As we will see when considering concrete examples, one can easily find a vector field $\Upsilon_\epsilon$, such that $\delta_\epsilon \mathfrak{f}=\delta f(\Upsilon_\epsilon)$. Hence, one can write
\begin{align}
    \delta_\epsilon \mathfrak{f}=\delta \mathfrak{f}(\Upsilon_\epsilon)= i_{\Upsilon_\epsilon} \Omega(X_\mathfrak{f})\stackrel{!}{=}\Set{\mathfrak{f}, \mathcal{Q}[\epsilon]}, \label{upsilondef}
\end{align}
where we used \eqref{phasespacevectorfielddef} and the antisymmetry of the symplectic form. The last equation is the requirement we put on the generating charge, i.e. it has to generate the symmetry variation. The bracket can also be rewritten as
\begin{align}
    \Set{\mathfrak{f}, \mathcal{Q}[\epsilon]}=i_{X_{\mathcal{Q}[\epsilon]}}\Omega(X_\mathfrak{f})=-\delta\mathcal{Q}[\epsilon](X_\mathfrak{f})\label{341}.
\end{align}
Comparing \eqref{upsilondef} to \eqref{341} and using that this has to hold for any $X_\mathfrak{f}$, we find the relation
\begin{align}
    \delta \mathcal{Q}[\epsilon]=-i_{\Upsilon_\epsilon}\Omega, \label{secondchargedefinition}
\end{align}
which also implies $\Upsilon_\epsilon=X_{\mathcal{Q}[\epsilon]}$ due to \eqref{phasespacevectorfielddef}.

\section{Electrodynamics}  
\label{Sec:ED}

In this section, we investigate asymptotic symmetries and memory effects for electrodynamics in decelerating and spatially flat FLRW spacetimes at future null infinity $\mathcal{I}$, as well as their interconnections. We begin by studying the asymptotic behaviour of scalar and Dirac field theories. These will be used as currents, motivating the fall-off conditions in our study of the asymptotic symmetries within electrodynamics. Finally, we explore the diverse associated memory effects.

\subsection{Currents}
\label{Subsec:currents}

In this subsection, we study the asymptotic behaviour of scalar and Dirac field theories for the decelerating and spatially flat FLRW background.

\subsubsection{Scalar field theory}
\label{Subsubsec:Scalar}

The main reason to study the behavior of the complex scalar field is to find the fall-off conditions for charge currents which we can then couple to the gauge fields. Apart from that, the behavior of scalar fields is of interest already on its own.

The action of a massless complex scalar field theory in curved spacetime is provided by
\begin{align}
S_\Phi=\int_{\mathcal{M}} \pd^4 x \sqrt{-g}  \left(g^{\mu\nu}\nabla_\mu \Phi^* \nabla_\nu \Phi +\xi R \Phi^* \Phi\right) \ , \label{conformallycoupledcomplexscalaraction}
\end{align}
where we introduced an extra term $\xi R \Phi^* \Phi$ with respect to the action in Minkowski spacetime ($\mathbb{M}$) to make the transition to curved spacetime (cf. \cite{BirrellDavies}).  $\xi\in \R$ is some coupling parameter and $R$ the Ricci scalar. From this action, we obtain the following equations of motion 
\begin{align}
\left(\Box-\xi R\right) \Phi =0 && \text{and} && \left(\Box -\xi R\right) \Phi^*=0.
\end{align}
The two standard choices for $\xi$ are the minimal coupling $\xi=0$ and the conformal coupling $\xi=\frac{1}{6}$. In the following, we will choose conformal coupling. The reason will become apparent in a moment. Since the action is invariant under complex conjugation, it suffices to examine $\Phi$ and transfer the results to $\Phi^*$. 
We begin by writing out the equations of motion in retarded Bondi coordinates
\begin{align}
\frac{1}{a^2}\left[\frac{2}{r} \frac{u+(s+1)r}{u+r}\p_u - \frac{2}{r}\p_r +2\p_u \p_r -\p_r^2-\frac{1}{r^2}D^2 + \frac{s(s-1)}{(u+r)^2}\right]\Phi=0. \label{scalarfieldeom}
\end{align}
Next, we introduce a rescaled field $\chi(u,r,\bth)\equiv a(u,r)\Phi(u,r,\bth)$ and plug it into \eqref{scalarfieldeom}
\begin{align}
\left[\frac{2}{r}(\p_u-\p_r)+2\p_u\p_r -\p_r^2-\frac{1}{r^2}D^2\right]\chi=\widehat{\Box}\chi=0, \label{conformallycoupledscalarfieldeom}
\end{align}
where $\widehat{\Box}$ is the d'Alembert operator on Minkowksi space. Hence, according to the equations of motion, $\chi$ behaves just like a free massless scalar field on $\mathbb{M}$, for which a treatment of the asymptotic behaviour can be found e.g. in \cite{MitraPhD}. We consider only the conformally coupled case because we want the charges to reach null infinity, meaning that they must be massless. For the minimally coupled scalar field, it is shown (e.g. in \cite{Winitzki}) that it behaves like effectively having a mass (which evolves in time) and we expect this to be the case for any coupling but conformal coupling \footnote{Let us note that if one applies the following analysis to the minimally coupled scalar field, one would find the same fall-off conditions. However, due to not being massless in the above sense, this result does not seem trustworthy, and we shall not pursue this case here.}.

In order to examine the behavior of $\chi$ near $\scri$, we take the limit $r\to \infty$ of \eqref{conformallycoupledscalarfieldeom}, assuming that $\chi$ can be expanded in an asymptotic series
\begin{align}
\left(\frac{1}{r}+\p_r\right) \p_u \chi^l=0. \label{conformallycoupledleadingorder}
\end{align}
Here, $\chi^l$ is the leading order term of the $u$-dependent part of $\chi$ in a large-$r$ expansion. We can unambiguously solve the equation by $\chi^l(u,r,\bth)=\frac{C_1(u,\bth)}{r}$ for some function $C_1$ on $\scri$, which motivates the following boundary condition
\begin{align}
\chi\in \mathcal{O}\left(r^{-1}\right) . \label{largerconditionscalarfield}
\end{align}
Note that we have implicitly neglected that \eqref{conformallycoupledleadingorder} puts no restrictions on the $u$-independent part of $\chi$. This is reasonable since we do not investigate the soft sector of scalar field theories and, hence, are not interested in the $u$-independent part of $\chi$.

From a physical point of view, we should also impose finiteness of energy, momentum and angular momentum flux through $\scri$ \footnote{Such a restriction, usually placed also for the Minkowski background, still makes sense in our context due to our requirement being only applied to the energy-momentum tensor of the gauge field, not the one of the FLRW background.}. To make this argument formal, the conserved current $T_{\mu\nu}^\mathbb{M}X^\nu$, where $X^\nu$ is the most general Killing vector on $\mathbb{M}$, is used in Minkowski space. However, since FLRW spacetimes are non-stationary, there is no timelike Killing vector and it would not make sense to require finite energy flux in general. In spite of this, the field theory we consider is conformally invariant. As a consequence, the stress-energy tensor of the conformally coupled scalar field is traceless. Therefore, we do not have to restrict ourselves to Killing vectors, but we can find a conserved current by replacing the Killing vector with a conformal Killing vector. A conformal Killing vector $X^\mu$ satisfies $\mathcal{L}_X g_{\mu\nu}=2\nabla_{(\mu}X_{\nu)}=\alpha g_{\mu\nu}$ for some scalar $\alpha$. 
$T_{\mu\nu}X^\nu$ is then a conserved quantity since 
\begin{align}
\nabla^\mu \left(T_{\mu\nu} X^\nu\right)= \nabla^\mu T_{\mu\nu} X^\nu + T_{\mu\nu} \nabla^{(\mu}X^{\nu)}= \frac{\alpha}{2} T_{\mu\nu} g^{\mu\nu}=0,
\end{align}
where we used in the first step that $T_{\mu\nu}=T_{(\mu\nu)}$, in the second step that $X^\nu$ is a conformal Killing vector and $\nabla^\mu T_{\mu\nu}=0$, and in the last step that the energy-momentum tensor is traceless.  

Due to the fact that spatially flat FLRW spacetimes and Minkowski space are conformally related, a Killing vector $X^\mu$ on $\mathbb{M}$ is a conformal Killing vector for $\mathcal{M}$. Moreover, since the stress-energy tensor satisfies $T_{\mu\nu}(\Phi)=\frac{1}{a^2}T_{\mu\nu}^{\mathbb{M}}(\chi)$ and we have for the integration over $\scri$ also the relation $\pd \Sigma^\mu=a^2\pd \Sigma^\mu_\mathbb{M}$, our condition simply becomes
\begin{align}
\left|\int_\scri T_{\mu\nu}(\Phi) X^\mu \pd \Sigma^\nu\right| = \left|\int_{\scri^+} T_{\mu\nu}^\mathbb{M}(\chi) X^\mu \pd \Sigma^\nu_\mathbb{M}\right|<\infty \ ,
\end{align}
which leads to the fall-off condition $\chi,\chi^*\in \sO(r^{-1})$, corresponding to the flat spacetime result that was found in \cite{HeMitraCov, MitraPhD}, leading to the conditions $\Phi,\Phi^*\in\sO\left((ar)^{-1}\right)$. 

More directly, we should also impose finite charge flux through $\scri$. The action \eqref{conformallycoupledcomplexscalaraction} is invariant under global $U(1)$-transformations, which leads for fixed background metric to the conserved Noether current 
\begin{align}
j_\mu=i\left(\nabla_\mu \Phi^* \Phi - \Phi^* \nabla_\mu \Phi\right).\label{conformallycouplednoethercurrent}    
\end{align}
Finiteness of charge flux means
\begin{align}
\left|\int_\scri j_\mu \pd \Sigma^\mu \right|=\left|i \int_\scri \lim_{r\to\infty} r^2 a^2(u,r) \left(\p_u \Phi^* \Phi - \frac{1}{2}\p_r \Phi^* \Phi\right)\pd u \pd\Gamma+ \text{c.c.}\right|<\infty.  
\end{align}
With the behavior $\Phi, \Phi^*\in \sO\left((ar)^{-1}\right)$, this is clearly satisfied and in agreement with the above conditions. Hence, we take the large-$r$ fall-offs to be
\begin{align}
\Phi\in\sO\left((ar)^{-1}\right).\label{scalarfieldfalloffs}
\end{align}
Using an asymptotic expansion of the form
\begin{empheq}{align}
\Phi(u,r,\bth)=\frac{1}{a(u,r)}\sum_{n=1}^\infty \frac{\Phi^{(n)}(u,\bth)}{r^n}, \label{scalarasymptoticexpansionFLRW}
\end{empheq}
we can consistently solve equation \eqref{scalarfieldeom}: For $n\geq 1$, we have
\begin{align}
2n \p_u \Phi^{(n+1)}=-n(n-1)\Phi^{(n)}-D^2 \Phi^{(n)}.
\end{align}
The corresponding result for the case of flat spacetime was found, e.g. in \cite{MitraPhD}. To solve this equation uniquely, we first need to completely determine $\Phi^{(1)}$. Then, we can find any higher order $\Phi^{(n+1)}$ upon specifying its value at a fixed $u$. Therefore, $\Phi^{(1)}$ is our boundary data.\\
For later purpose, notice that when inserting the asymptotic expansion of the scalar field into \eqref{conformallycouplednoethercurrent}, we obtain
\begin{align}
    j_\mu(u,r,\bth)=\frac{j_\mu^\mathbb{M}(u,r,\bth)}{a^2(u,r)}, \label{conformallycoupledcurrent}
\end{align}
where $j_\mu^\mathbb{M}=i\left(\p_\mu \chi^*\chi-\chi^*\p_\mu \chi\right)$.\\
Let us also construct the corresponding matter phase space on $\scri$. From the variation of  \eqref{conformallycoupledcomplexscalaraction}, we find the following symplectic potential and current
\begin{align}
    \Theta_\mu= \nabla_\mu \Phi \delta \Phi^* + \text{c.c.} && \text{and} && \omega_\mu=\nabla_\mu \delta \Phi \curlywedge \delta \Phi^* +\text{c.c.},
\end{align}
and using the radial fall-off conditions \eqref{scalarfieldfalloffs}, we get for the symplectic form
\begin{align}
    \Omega_{\scri}^\Phi=\int_{\scri} \p_u \delta \Phi^{(1)}(u,\bth) \curlywedge \delta \Phi^{(1)*}(u,\bth) \pd u \wedge \pd \Gamma +\text{c.c.}
\end{align}
Since we do not investigate the soft sector of scalar field theories, we assume in the leading radial order of the scalar field the additional condition $\lim_{u\to\pm\infty}\Phi^{(1)}(u,\bth)=0$. The symplectic form is then indeed invertible and we obtain the only non-vanishing Poisson bracket as
\begin{align}
    \left\{\Phi^{(1)}(u, \bth), \Phi^{(1)*}(u',\bth')\right\}=\frac{1}{4}\sgn(u-u')\delta_{S^2}(\bth, \bth').
\end{align}
Taking the flat spacetime limit of our results, one obtains the brackets found for a Minkowski background in \cite{MitraPhD}.

\subsubsection{Dirac field theory}
\label{Subsubsec:Dirac}

In the following, we investigate the asymptotic behavior of spin-$\frac{1}{2}$ fields for FLRW spacetimes. This analysis is interesting in itself and has already fallen short in the literature in the case of a Minkowski background. The only analysis we are aware of which considers spin-$\frac{1}{2}$ fields (in flat spacetime and only left-handed ones) is \cite{MitraPhD}. Our results are consistent with those. Moreover, Dirac fields can serve as another form of matter we can couple to our gauge fields. 

The action of a spin-$\frac{1}{2}$ particle described by a 4-component Dirac spinor is
\begin{align}
S_\Psi=\frac{i}{2}\int_{\mathcal{M}} \sqrt{-g}\pd^4 x \Bar{\Psi}\gamma^\mu \nabla_\mu \Psi + \text{h.c.} \label{spinonehalfaction}
\end{align}

The equations of motion are given by
\begin{align}
    i\gamma^\mu \nabla_\mu \Psi=0 && \text{and} && i\left(\nabla_\mu \Bar{\Psi}\right) \gamma^\mu=0.
    \label{eq:diraceqn}
\end{align}

Analogously to the complex scalar field, we restrict our attention to $\Psi$ since $\Bar{\Psi}$ must have the same fall-off behavior. Since we consider herein massless fields, we have chosen the chiral basis for the gamma matrices, which allows to split up the equations of motion by introducing $\Psi=(\Psi_L, \Psi_R)^T$ where $\Psi_{L,R}$ are Weyl-spinors. In retarded Bondi coordinates \eqref{eq:diraceqn} takes, under application of the conventions in appendix \ref{Appendix:Spinorconventions}, the form
\begin{subequations}
\begin{align}
    \frac{1}{a}\begin{pmatrix} \p_u +\frac{3s}{2(u+r)}+\frac{\p_\varphi}{r\sin(\theta)} & \p_r-\p_u +\frac{1}{r}\left(1-i\p_\theta-\frac{i}{2\tan(\theta)}\right) \\ \p_r-\p_u +\frac{1}{r}\left(1+i\p_\theta+\frac{i}{2\tan(\theta)}\right) & \p_u +\frac{3s}{2(u+r)}-\frac{\p_\varphi}{r\sin(\theta)}
    \end{pmatrix}\Psi_R=0,\\
    \frac{1}{a}\begin{pmatrix}\p_u + \frac{3s}{2(u+r)}-\frac{\p_\varphi}{r\sin(\theta)} & \p_u-\p_r-\frac{1}{r}\left(1-i\p_\theta-\frac{i}{2\tan(\theta)}\right)\\ \p_u-\p_r-\frac{1}{r}\left(1+i\p_\theta+\frac{i}{2\tan(\theta)}\right) & \p_u+\frac{3s}{2(u+r)}+\frac{\p_\varphi}{r\sin(\theta)}
    \end{pmatrix}\Psi_L=0.
\end{align}
\end{subequations}
Noting that $\p_u a^{-3/2}=\p_r a^{-3/2}=-\frac{1}{a^{3/2}} \frac{3s}{2(u+r)}$, we can simplify these equations by introducing an auxiliary field $\zeta(u,r,\bth)\equiv a^{3/2}(u,r)\Psi(u,r,\bth)$:
\begin{subequations}
\begin{align}
    \begin{pmatrix} \p_u +\frac{\p_\varphi}{r\sin(\theta)} & \p_r-\p_u +\frac{1}{r}\left(1-i\p_\theta-\frac{i}{2\tan(\theta)}\right) \\ \p_r-\p_u +\frac{1}{r}\left(1+i\p_\theta+\frac{i}{2\tan(\theta)}\right) & \p_u -\frac{\p_\varphi}{r\sin(\theta)}
    \end{pmatrix}\zeta_R=0,\\
    \begin{pmatrix}\p_u -\frac{\p_\varphi}{r\sin(\theta)} & \p_u-\p_r-\frac{1}{r}\left(1-i\p_\theta-\frac{i}{2\tan(\theta)}\right)\\ \p_u-\p_r-\frac{1}{r}\left(1+i\p_\theta+\frac{i}{2\tan(\theta)}\right) & \p_u+\frac{\p_\varphi}{r\sin(\theta)}
    \end{pmatrix}\zeta_L=0.
\end{align}
\end{subequations}
This is clearly the same result we would have obtained by setting $s=0$ which corresponds to a scale factor $a=1$. Hence, $\zeta$ behaves like a Dirac field on a Minkowski background. 
For the following analysis, it turns out to be beneficial to rotate the basis from $\zeta_{L,R}=\left(\zeta_{L,R}^1, \zeta_{L,R}^2\right)$ according to
\begin{align}\label{plusminusspinor}
    \zeta_L^\pm=\frac{1}{\sqrt{2}}\left(\zeta_L^1\mp\zeta_L^2\right), && \zeta_R^\pm=\frac{1}{\sqrt{2}}\left(\zeta_R^1\pm\zeta_R^2\right),
\end{align}

by means of which the equations of motion become

\begin{subequations}\begin{align}    \left[2\p_u-\p_r\right]\left(r\zeta_R^-\right)+\left[\frac{\p_\varphi}{\sin(\theta)}-i\p_\theta-\frac{i}{2\tan(\theta)}\right]\zeta_R^+=0, \label{Dirac1} \\
\left[\frac{\p_\varphi}{\sin(\theta)}+i\p_\theta+\frac{i}{2\tan(\theta)}\right]\zeta_R^-+\p_r\left(r\zeta_R^+\right)=0\label{Dirac2},\\    \p_r\left(r\zeta_L^+\right)-\left[\frac{\p_\varphi}{\sin(\theta)}-i\p_\theta-\frac{i}{2\tan(\theta)}\right]\zeta_L^-=0, \label{Dirac3}\\    \left[\frac{\p_\varphi}{\sin(\theta)}+i\p_\theta+\frac{i}{2\tan(\theta)}\right]\zeta_L^+-\left[2\p_u-\p_r\right]\left(r\zeta_L^-\right)=0. \label{Dirac4}\end{align}
\end{subequations}
Assuming that $\zeta$ can be expanded in terms of $\frac{1}{r}$ in the limit $r\to \infty$, we conclude from \eqref{Dirac1} that $\zeta_R^-\in \sO\left(r^{-1}\zeta_R^+\right)$. Equation \eqref{Dirac2} then implies that $\zeta_R^+\in \sO(r^{-1})$ and hence $\zeta_R^-\in \sO(r^{-2})$.  In a similar fashion, we infer from equations \eqref{Dirac3} and \eqref{Dirac4} that $\zeta_L^+\in \sO(r^{-1})$ and $\zeta_{L}^-\in \sO(r^{-2})$.

Let us check that these conditions guarantee the finiteness of charge, energy, momentum and angular momentum flux through $\scri$. For the charge flux, we first construct the Noether current associated with the global $U(1)$-symmetry of the theory which reads (cf. appendix \ref{Appendix:Spinorconventions})
\begin{align}
    j_\mu=\Bar{\Psi}\gamma_\mu \Psi=\Psi_R^\dagger \sigma_\mu \Psi_R+ \Psi_L^\dagger \Bar{\sigma}_\mu \Psi_L.
\end{align}
In Bondi coordinates, we find
\begin{subequations}\label{Diracfieldchargecurrent}
\begin{align}
j_u=a\left(\Psi_R^{+*}\Psi_R^++\Psi_R^{-*}\Psi_R^-+\Psi_L^{+*}\Psi_L^++\Psi_L^{-*}\Psi_L^-\right)\in\sO\left(a^{-2}r^{-2}\right), \\
    j_r=2a\left(\Psi_R^{-*}\Psi_R^-+\Psi_L^{-*}\Psi_L^-\right)\in\sO\left(a^{-2}r^{-4}\right),\\
    j_\theta=iar\left(\Psi_R^{-*}\Psi_R^+-\Psi_R^{+*}\Psi_R^--\Psi_L^{+*}\Psi_L^-+\Psi_L^{-*}\Psi_L^+\right)\in\sO\left(a^{-2}r^{-2}\right),\\
    j_\varphi=ar\sin(\theta)\left(\Psi_L^{-*}\Psi_L^++\Psi_L^{+*}\Psi_L^--\Psi_R^{-*}\Psi_R^+-\Psi_R^{+*}\Psi_R^-\right)\in\sO\left(a^{-2}r^{-2}\right) .
\end{align}
\end{subequations}
The finite charge flux condition merely imposes the fall-offs $j_u,j_r\in\sO(a^{-2}r^{-2})$ and is clearly satisfied.

Due to the conformal invariance of the theory, we can apply the same reasoning we used for the scalar field in the paragraph below \eqref{largerconditionscalarfield} to argue that our second finiteness condition takes the form $\left|\int_\scri T_{\mu\nu}(\Psi) X^\mu \pd \Sigma^\nu\right|=\left|\int_{\scri^+} T_{\mu\nu}^\mathbb{M}(\zeta) X^\mu \pd \Sigma^\nu_{\mathbb{M}}\right|<\infty$.\\
Therefore, the suggested asymptotic expansion for the physical field reads
\begin{align}
    \Psi_{L,R}^+(u,r,\bth)=\frac{1}{a^{\frac{3}{2}}(u,r)}\sum_{n=1}^\infty \frac{\Psi_{L,R}^{+,(n)}(u,\bth)}{r^n}, && \Psi_{L,R}^{-}(u,r,\bth)=\frac{1}{a^{\frac{3}{2}}(u,r)}\sum_{n=2}^\infty \frac{\Psi_{L,R}^{-,(n)}(u,\bth)}{r^n}. \label{Diracexpansion}
\end{align}
Plugging this expansion into the equations of motion, we obtain for $n\geq 1$
\begin{subequations}\label{Diracexpansionequations}
\begin{align}
    2\p_u \Psi_R^{-,(n+1)}+(n-1)\Psi_R^{-,(n)}+\left[\frac{\p_\varphi}{\sin(\theta)}-i\p_\theta-\frac{i}{2\tan(\theta)}\right]\Psi_R^{+,(n)}=0,\label{Diracexpansion1}\\
    \left[\frac{\p_\varphi}{\sin(\theta)}+i\p_\theta+\frac{i}{\tan(\theta)}\right] \Psi_R^{-,(n+1)}-n\Psi_R^{+,(n+1)}=0,\label{Diracexpansion2}\\
    n\Psi_L^{+,(n+1)}+\left[\frac{\p_\varphi}{\sin(\theta)}-i\p_\theta-\frac{i}{\tan(\theta)}\right]\Psi_L^{-,(n+1)}=0,\label{Diracexpansion3}\\
    \left[\frac{\p_\varphi}{\sin(\theta)}+i\p_\theta+\frac{i}{2\tan(\theta)}\right]\Psi_L^{+,(n)}-2\p_u\Psi_L^{-,(n+1)}-(n-1)\Psi_L^{-,(n)}=0. \label{Diracexpansion4}
\end{align}
\end{subequations}
In the case of flat spacetime, analogous equations were found in \cite{MitraPhD} for the left-handed spinor field. We can solve these equations iteratively. For the right-handed part, specifying $\Psi_R^{+,(1)}$ completely, we are due to \eqref{Diracexpansion1} able to determine $\Psi_R^{-,(2)}$ completely upon fixing its value at a specific $u$. Using then \eqref{Diracexpansion2} and \eqref{Diracexpansion1} alternately and specifying for any $n$ the values of $\Psi_R^{-,(n)}$ at a certain value of $u$, we can solve the equations uniquely. In the same manner, we can determine the left-handed part uniquely by specifying $\Psi_L^{+,(1)}$ completely and fixing the value of $\Psi_R^{-,(n)}$ at a given $u$ for any $n$. Therefore, $\Psi_L^{+,(1)}$ and $\Psi_R^{+,(1)}$ are our boundary data on $\scri$. 

Finally, let us find the corresponding phase space on $\scri$. From the variation of the action \eqref{spinonehalfaction}, we obtain the symplectic potential and current as
\begin{align}
    \Theta_\mu=\frac{i}{2}\left(\Bar{\Psi}\gamma_\mu \delta\Psi -\delta \Bar{\Psi}\gamma^\mu \Psi\right) && \text{and} && \omega_\mu=i\delta\Bar{\Psi}\curlywedge\left( \gamma_\mu\delta \Psi\right)  . \end{align}
Note that in this case, the exterior product on the phase space must be symmetric and not antisymmetric, i.e. $\delta \Psi_1 \curlywedge \delta \Psi_2=\delta \Psi_2 \curlywedge\delta \Psi_1$,
since we are dealing with fermions instead of bosons now. Correspondingly, we will in the end obtain symmetric brackets instead of Poisson brackets such that in a canonical quantization we would replace them by anticommutators rather than commutators.\footnote{We will comment on this and associated problems in more detail in section \ref{Sec:YM}. See also \cite{grassmanngraded}.} \\
Using the expansion \eqref{Diracexpansion}, one finds for the symplectic form \footnote{Note that this is a generalization of a symplectic form.} on $\scri$ to leading order
\begin{align}
    \Omega_{\scri}^\Psi=i\lim_{r\to\infty}\int_\scri a^2 r^2 \delta \Bar{\Psi}\curlywedge\left(\left(\gamma_u-\frac{1}{2}\gamma_r\right)\delta \Psi\right) \pd u\wedge \pd \Gamma=\nonumber \\=i\int_\scri \left(\left(\delta\Psi_R^{+,(1)}\right)^*\curlywedge \delta\Psi_R^{+,(1)}+\left(\delta\Psi_L^{+,(1)}\right)^*\curlywedge \delta\Psi_L^{+,(1)}\right)\pd u\wedge\pd\Gamma. \label{eq:DiracSpinor_SymplecticForm}
\end{align}
It is invertible and we can read off the only non-vanishing symmetric brackets as
\begin{subequations}
\begin{align}
    \left\{{\Psi_R^{+,(1)}}^*(u,\bth), \Psi_R^{+,(1)}(u',\bth')\right\}=i\delta(u-u')\delta_{S^2}(\bth,\bth'), \\ \left\{{\Psi_L^{+,(1)}}^*(u,\bth), \Psi_L^{+,(1)}(u',\bth')\right\}=i\delta(u-u')\delta_{S^2}(\bth,\bth'). \label{eq:DiracSpinor_LeftBracket}
\end{align}
\end{subequations}

In the flat limit, our results for $\Psi_L$ are compatible with those found in \cite{MitraPhD} until equation \eqref{eq:DiracSpinor_SymplecticForm}. The associated bracket \eqref{eq:DiracSpinor_LeftBracket} shows in our case, however, no mixing between $\Psi_L^{+,(1)}$ and $\Psi_L^{-,(2)*}$, contrarily to suggested in \cite{MitraPhD}. An argument in favor of the brackets we derived is that we showed below \eqref{Diracexpansionequations} that $\Psi_L^{+,(1)}$ and its complex conjugate are the boundary data on $\scri$ and, consequently, $\Psi_L^{-,(2)}$ should not be present on the phase space. The inclusion of $\Psi_L^{-,(2)}$ (and its complex conjugate) on the phase space would suggest the presence of four degrees of freedom for the left-handed Weyl spinor, even though it is well known that it carries only two.

\subsection{Asymptotic symmetries}
\label{Subsec:asympsym}

All the necessary tools to deal with asymptotic symmetries in decelerating and spatially flat FLRW are now available. Due to the conformal invariance of the pure Maxwell action, it turns out that the procedure and results are formally identical to those in $\mathbb{M}$. Nevertheless, we perform a detailed derivation and use this opportunity to point out a major inconsistency, already present in the flat space analysis, which has gone largely unnoticed in the literature.

\vspace{1em}

The action of electrodynamics is given by
\begin{align}
S_\text{ED}=-\int_{\sM} \pd^4 x \sqrt{-g} \left\{\frac{1}{4}F_{\mu\nu} F^{\mu\nu}+A_\mu J^\mu \right\},   \label{freeEDactionFLRW}
\end{align}
where $J_\mu$ is a non-dynamical, gauge invariant, and conserved current.
Motivated by the fall-offs of the matter currents studied in section \ref{Subsec:currents}, we impose the conditions
\begin{align}
    J_u\in \sO(a^{-2}r^{-2}), && J_r\in \sO(a^{-2}r^{-4}), && J_B\in \sO(a^{-2}r^{-2}). \label{FLRWcurrentfalloff}
\end{align}
The action \eqref{freeEDactionFLRW} is conformally invariant upon rescaling the charge current by $a^2$. In this sense, the equations of motion $\nabla^\mu F_{\mu\nu}=J_\nu$ can for example be written as $\widehat{\nabla}^\mu F_{\mu \nu}=a^2 J_\nu$, where $\widehat{\nabla}^\mu$ is the covariant derivative on $\mathbb{M}$. Due to the conformal invariance,
we can apply the same reasoning we used for the scalar field in the paragraph after \eqref{largerconditionscalarfield} to conditions like the finiteness of energy, momentum and angular momentum flux through $\scri$, to argue that $\left|\int_\scri T_{\mu\nu}(F) X^\mu \pd \Sigma^\nu\right|=\left|\int_{\scri^+} T_{\mu\nu}(F) X^\mu \pd \Sigma^\nu_{\mathbb{M}}\right|<\infty$. 
We can therefore apply a similar analysis as it has been developed for flat spacetime in the literature. 
These conditions impose, together with the equations of motion and the matter current fall-offs from the previous subsections, the following asymptotic behaviour of the field strength tensor 
\begin{align}
    F_{ur}\in \sO(r^{-2}), && F_{uA}\in \sO(r^0), && F_{rA}\in \sO(r^{-2}), && F_{AB}\in \sO(r^{0}), \label{fieldstrengthfalloffsFLRW}\\
    && \lim_{u\to\pm\infty} \mathcal{F}_{uB}=0, && && \mathcal{F}_{AB}=0. \label{magneticabsenceenergyfinite}
\end{align}
The last two constraints come from the finiteness of energy flux and the absence of magnetic charges, respectively. Here and in the following, calligraphic letters denote the leading order in an asymptotic expansion, e.g., $\mathcal{F}_{ur}(u,\bth)=\lim_{r\to\infty}\left(r^2 F_{ur}(u,r,\bth)\right)$.
Next, we fix the retarded radial gauge $A_r=0$ and $A_u\big\vert_{\scri}=0$. In this gauge, it is furthermore possible to apply an asymptotic expansion and radial fall-off conditions of the form 
\begin{align}
    J_\mu(u,r,\bth)=\sum_{n=0}^\infty \frac{J^{(n)}_\mu(u,\bth)}{a^2(u,r) r^n} && A_\mu(u,r,\bth)=\sum_{n=0}^\infty \frac{A^{(n)}_\mu (u,\bth)}{r^n}, \label{gaugefieldexpansionFLRW}
\end{align}
\vspace{-0.5cm}
\begin{align}
    A_u\in \sO(r^{-1}), && A_r=0, &&  A_B\in \sO(r^0). \label{EDfalloffconditionsFLRW}
\end{align}
This is consistent with the equations of motion and the fall-off conditions for the field strength \eqref{fieldstrengthfalloffsFLRW}.

\vspace{1em}

Our next objective is to construct the phase space on $\scri$ and find the fundamental Poisson brackets following the procedure laid out in section \ref{Sec:CPS}. From the boundary term of the variation of the action \eqref{freeEDactionFLRW}, we can read off the symplectic potential and, hence, the symplectic current as
\begin{align}
\Theta_\mu =-F_{\mu\nu} \delta A^\nu && \text{and} && \omega_\mu=-(\delta(\p_\mu A_\nu)-\delta(\p_\nu A_\mu))\curlywedge \delta A^\nu.
\end{align}
The presymplectic form on $\scri$ can then be found by appyling the above fall-off conditions for the gauge field:
\begin{align}
\tilde{\Omega}_{\scri}=\int_{\scri} \omega_\mu \pd \Sigma^\mu=\int_{\scri} \gamma^{AB} \delta (\p_u \cA_A)\curlywedge \delta \cA_B \pd u \wedge \pd \Gamma.\label{freeEDsymplecticform} 
\end{align}
If this presymplectic form were invertible, we would call it the symplectic form $\tilde{\Omega}_{\scri}=\Omega_{\scri}$ and obtain Poisson brackets between $\mathfrak{f}, \mathfrak{g}\in\mathfrak{F}(\Gamma)$ as $\Set{\mathfrak{f},\mathfrak{g}}=-\Omega_{\scri}(X_\mathfrak{f}, X_\mathfrak{g})$ (cf. section \ref{Sec:CPS}). 
To obtain the fundamental brackets we would have to choose $\mathfrak{f}=\cA_C(u,\bth)$ and $\mathfrak{g}=\cA_D(u', \bth')$. In order to determine $X_\mathfrak{f}$, we begin by writing the general form of a vector field on phase space
\begin{align}
X_\mathfrak{f}=\int_{\scri} x_{\mathfrak{f};C}(u,\bth) \frac{\delta}{\delta \cA_C(u,\bth)} \pd u\wedge \pd \Gamma. 
\end{align}
Making use of it, we find upon integration by parts
\begin{align}
i_{X_\mathfrak{f}} \Omega_{\scri}=2\int_{\scri}  \gamma^{AB} \p_u x_{\mathfrak{f};A}(u,\bth)\delta \cA_B(u,\bth)\pd u\wedge \pd \Gamma -\oint_{S^2}  \gamma^{AB} x_{\mathfrak{f};A}(u,\bth)\delta \cA_B(u,\bth)\Big\vert_{u=\pm \infty}\pd \Gamma.
\end{align}
Since $X_\mathfrak{f}$ has to satisfy by definition $i_{X_\mathfrak{f}}\Omega_{\scri}=-\delta \mathfrak{f}$, the above expression has to be the same as
$-\delta \mathfrak{f}=-\delta \cA_C(u,\bth)=-\int_\scri \gamma^{AB} \gamma_{AC} \delta(u-\tilde{u})\delta_{S^2}(\bth, \tilde{\bth}) \delta \cA_B(\tilde{u}, \tilde{\bth})\pd\tilde{u}\wedge \tilde{\bth}$. This has to be true for arbitrary $\delta \cA_C$, in particular for those which are non-vanishing at $u=\pm\infty$, and thus implies $x_{\mathfrak{f};A}(\pm \infty,\bth)=0$ and so $\gamma_{AC} \delta(u-\tilde{u})\delta_{S^2}(\bth, \tilde{\bth})=-2\p_{\tilde{u}} x_{\mathfrak{f};A}(\tilde{u},\tilde{\bth}) $. Integrating this equation over $\tilde{u}$ would then, however, lead to the contradiction $\gamma_{AC}\delta_{S^2}(\bth, \tilde{\bth})=0$. Therefore, the Poisson brackets we are looking for are not properly defined. The same is shown in \cite{Mohd, CampigliaLaddha}, applying the above procedure to toy models. The problem arises due to the phase space being infinite dimensional. While the symplectic form we found is non-degenerate \footnote{As long as we do not fix the boundary values such that $\cA_B\big\vert_{u=\infty}=\cA_B\big\vert_{u=-\infty}$.}, it is not invertible.\\ The problem can be resolved by extending the phase space, which was performed in the case of electromagnetism originally in \cite{NewSymmetriesofMasslessQED}. 
The first step towards this extension is to separate out the $u$-independent part of the gauge field $G_B(\bth)=\frac{1}{2}\left(\cA_B^+(\bth)+\cA_B^-(\bth)\right)$ such that
$\cA_B(u,\bth)=\hat{\cA}_B(u,\bth)+G_B(\bth)$. In particular, this implies
\begin{align}
    \hat{\cA}_B^+(\bth)=-\hat{\cA}_B^-(\bth). \label{uboundaryahat}
\end{align}
We use henceforth upper indices ``$\pm$" on the fields to denote the limits $u\to\pm \infty$, respectively, e.g., $\cA_B^+(\bth)=\lim_{u\to\infty} \cA_B(u,\bth)$. Rewriting $\cA_B$ in terms of $\hat{\cA}_B$ and $G_B$, the symplectic form reads 
\begin{align}
\Omega_{\scri} = \int_{\scri} \gamma^{AB} \delta (\p_u \hat{\cA}_A(u,\bth))\curlywedge \delta \hat{\cA}_B(u,\bth) \pd u \wedge \pd \Gamma + \oint_{S^2} \gamma^{AB} \delta N_A(\bth)\curlywedge \delta G_B(\bth) \pd \Gamma,\label{symplecticform}
\end{align}
where we have furthermore defined $N_B(\bth)=\cA_B^+(\bth)-\cA_B^-(\bth)$. Rewriting $\Omega_\scri$ in this way does not introduce any change concerning its invertibility. The crucial step lies in considering the gauge stripped field $\hat{\cA}_B$ and the two $u$-independent fields $N_B$ and $G_B$ as independent of each other. \footnote{In this context, independent means that the derivatives of the fields with respect to each other vanish. In other words, they are independent coordinates on the phase space.}

Recall that, due to the absence of magnetic charges, $\mathcal{F}_{AB}=0$ and, hence, we can write $\cA_B^\pm(\bth)=\p_B\alpha^\pm(\bth)$. Defining $N\equiv \alpha^+-\alpha^-$ and $G\equiv\frac{1}{2}\left(\alpha^++\alpha^-\right)$, such that $N_B=\p_B N$ and $G_B=\p_B G$, the symplectic form becomes
\begin{align}
\Omega_{\scri} = \int_{\scri} \gamma^{AB} \delta (\p_u \hat{\cA}_A(u,\bth))\curlywedge \delta \hat{\cA}_B(u,\bth) \pd u \wedge \pd \Gamma + \oint_{S^2} \gamma^{AB} \delta (D_A N(\bth)) \curlywedge \delta(D_B G(\bth)) \pd \Gamma. \label{Minksymplecticfinal}
\end{align}
To make this form non-degenerate, we have to introduce an equivalence relation $N\sim \Tilde{N} \Longleftrightarrow \exists n: N=\tilde{N}+n$ and $D^2n=0$. Since $n$ is a function on the sphere, the latter is just equivalent to $n$ being a constant. Exactly the same follows for $G$.

It is then indeed possible to extract from the above symplectic form the following non-vanishing fundamental Poisson brackets \footnote{In parts of the literature, Poisson brackets between $G$ and $N$ instead of $\p_A G$ and $\p_B N$ are written down. As noted in \cite{Miller}, such brackets do not exist as $G$ and $N$ are ambiguously defined.} 
\begin{subequations}
\label{PoissonbracketsMink}
\begin{align}
\Set{\hat{\cA}_A(u,\bth), \hat{\cA}_B(u',\bth')}&=-\frac{\gamma_{AB}}{4}\sgn(u-u')\delta_{S^2}(\bth,\bth'), \label{PoissonbracketsMinkgaugefield}\\
\Set{\p_A G(\bth), \p_B N(\bth')}&=\gamma_{AB}\delta_{S^2}(\bth, \bth'). \label{boundarybracketsMinkED}
\end{align}
\end{subequations}
In order to obtain the former, it is crucial to use the relation $\delta\hat{\cA}_B^+(\bth)=-\delta\hat{\cA}_B^-(\bth)$ following from \eqref{uboundaryahat}. The difficulties we faced in finding the Poisson brackets, which lead to the extension of phase space, are not specific to electrodynamics. They also occur for scalar field theory or more general Yang-Mills theories in the moment when one assumes that the respective fields do not vanish at the boundaries of $\scri$. The reason for these issues is rather that we integrated over a null hypersurface instead of a spacelike hypersurface, which we are more used to. Consider for example an equal-time slice. There, one would have obtained a pairing between the fields and their time-derivatives (conjugate momenta) in the symplectic form, which are on an equal time-slice clearly unrelated. Our analysis is, however, situated on an equal $v$-hypersurface. In this case, we obtained a symplectic pairing between the fields and their $u$-derivatives. Since $u$ is the coordinate running along the constant $v$-hypersurface, the field and its $u$-derivative are of course not independent which causes such difficulties.

\vspace{1em}

Previously, we fixed the retarded radial gauge. This leaves residual gauge transformations which act only on the leading order angular part of the gauge field in a non-trivial way, namely as
\begin{align}
\cA_B(u,\bth)\longrightarrow \cA_B(u,\bth)+\p_B \epsilon(\bth) . \label{freeEDresidualgauge}
\end{align}
For the fields on the extended phase space, this implies the transformation behaviour
\begin{align} \label{transformationbehaviourED}
    \hat{\cA}_B(u,\bth)\longrightarrow \hat{\cA}_B(u,\bth), &&
    N_B(\bth)\longrightarrow N_B(\bth), &&
    G_B(\bth)\longrightarrow G_B(\bth)+\p_B\epsilon(\bth).
\end{align}
Application of Noether's procedure allows one to find a conserved current $j_\mu\equiv\epsilon J_\mu+ \frac{\gamma^{AB}}{a^2 r^2} F_{\mu A}\p_B \epsilon$ associated with the above symmetry. Thus, we can construct the corresponding asymptotic charges under usage of the asymptotic fall-offs
\begin{align}
\mathcal{Q}[\epsilon]=\int_\scri j_\mu \pd \Sigma^\mu= \int_{\scri} \mathcal{J}_u(u,\bth) \epsilon(\bth)  \pd u \wedge \pd \Gamma+\int_{S^2} \gamma^{AB} N_A(\bth) \p_B \epsilon(\bth)  \pd \Gamma.\label{FLRWEDgeneratingcharge}
\end{align}
Note that, in contrast to the charge defined on $\scri^+$ for Minkowski space (cf. \cite{tobias:2022}), we have for FLRW spacetimes $\mathcal{J}_u=\lim_{r\to\infty} a^2 r^2 J_u$.

With the Poisson brackets derived in \eqref{PoissonbracketsMink}, we can see that $\mathcal{Q}[\epsilon]$ indeed generates the asymptotic symmetries. Explicitly, we find
\begin{subequations}
\begin{align}
    \Set{\hat{\cA}_B(u,\bth),\mathcal{Q}[\epsilon]}&=0,\\
    \Set{N_B(\bth), \mathcal{Q}[\epsilon]}&=0, \\
    \Set{G_B(\bth), \mathcal{Q}[\epsilon]}=\oint_{S^2} \gamma^{CD} \p_C \epsilon(\bth')\underbrace{\Set{G_B(\bth), N_D(\bth')}}_{=\gamma_{BD}\delta_{S^2}(\bth,\bth')}\pd \Gamma'&=\p_B\epsilon(\bth)
\end{align}
\end{subequations}
in agreement with \eqref{transformationbehaviourED}. As the soft photon fields commute with each other, it is also not hard to see that the charges satisfy the trivial $U(1)$-algebra
\begin{align}
\Set{\mathcal{Q}[\epsilon], \mathcal{Q}[\epsilon']}=0. \label{U1chargealgebra}
\end{align}
A naive approach points to the procedure for constructing the phase space being successful. However, at this point, we would like to draw the attention of the reader to a serious problem. In order to make the symplectic form invertible, we had to assume that the fields $\hat{\cA}_B, N_B$ and $G_B$ are independent of each other and, thus, had to forget about the original definition of the soft photon field $N_B=\int_\R \p_u \cA_B \pd u=\int_\R\p_u \hat{\cA}_B\pd u$. 
Unfortunately, this definition is necessary to relate asymptotic symmetries to memory effects and, in flat spacetime, to soft theorems. 
As one might have guessed from the derivation of the extended phase space, this definition is inconsistent with the Poisson brackets, as we will explicitly demonstrate now. One of the vanishing Poisson brackets was $\Set{G_A(\bth), \hat{\cA}_B(u,\bth')}=0$ for any $u$. Therefore, it clearly follows that
\begin{align}
    \Set{G_A(\bth), N_B(\bth')}=\int_\R \pd u \p_u \Set{G_A(\bth), \hat{\cA}_B(u,\bth')}=0 , 
\end{align}
where we used the linearity of the Poisson brackets. Nevertheless, this is in contradiction with \eqref{boundarybracketsMinkED}. To our knowledge, this inconsistency has remained unnoticed until the present work. Future studies should attempt to address this fundamental problem.

\subsection{Memory effects}
\label{Subsec:Memories}

This section focuses on the effect of an expanding universe on the relation between asymptotic symmetries and memory effects. Electromagnetic memory effects for cosmological backgrounds have been, to our knowledge, only considered in \cite{Chu:2016ngc}. However, therein these were only investigated under the assumption of short time scales over which the scale factor can be treated as a constant and without relating them to asymptotic symmetries. On the other hand, \cite{Chu:2016ngc} contains a more detailed treatment of radiative solutions of Maxwell's equations, making it thus complementary to the present analysis. 

To obtain the memory effect, we investigate how a non-vanishing difference between initial and final radiative vacuum is related to the motion of a test particle of mass $m$ and electric charge $q$, far away from all radiation sources. Defining its four-momentum as $p^\mu=m\frac{\pd x^\mu}{\pd \tau}$ with proper time $\tau$, the Lorentz force acting on the test particle in a curved background is given by 
\begin{align}
    \frac{\text{D}p^\alpha}{\pd \tau}=\frac{\pd p^\alpha}{\pd \tau} +\frac{1}{m}\Gamma^{\alpha}_{\mu\nu} p^\mu p^\nu=\frac{q}{m}g^{\alpha\beta}F_{\beta\gamma} p^\gamma.
\end{align}
We are going to assume now that the charged test particle is almost at rest with respect to the cosmic fluid. Explicitly, this means that $|p^r|, r|p^A|\ll p^u$, such that 
\begin{align}
m^2=p_\mu p^\mu=a^2\left((p^u)^2+2p^u p^r -r^2 \gamma_{AB} p^A p^B\right)\simeq a^2 (p^u)^2,
\end{align}
resulting in $p^u\simeq \frac{m}{a}$. Expanding the above Lorentz force equation to zeroth order in $p^r/m$ and $r p^A/m$, and, furthermore, assuming that our particle is at large enough $r$ to consider only the radiative part of the fields (cf. \eqref{fieldstrengthfalloffsFLRW}), we find for the change in momentum tangential to the large sphere
\begin{align}
\frac{\pd p^A}{\pd \tau}+\frac{2}{a}\Gamma^{A}_{uB} p^B\simeq  \frac{q\gamma^{AB}}{a^3 r^2} \p_u \cA_B  ,
\end{align}
where we used $\mathcal{F}_{uB}=\p_u \cA_B$ and $\mathcal{F}_{AB}=0$ due to the absence of magnetic charges. The term containing $\mathcal{F}_{AB}$ would have dropped out anyway to leading order of the slow motion approximation. Since the memory effect is the part of the asymptotic structure which is related to potential experiments, it makes sense to express the slow-motion approximation in terms of more observation-related quantities. We introduce the (rescaled) peculiar velocity as $\bv=a\frac{\pd \bx}{\pd t}$, where $\bx$ is the comoving coordinate with respect to the cosmic fluid. We can then write $\frac{\pd \tau}{\pd t}=\sqrt{1-\bv^2}\simeq 1$ and, hence, $\pd \tau\simeq a \pd \eta\simeq a \pd u$. As a consequence, the previous equation becomes
\begin{align}
    a^2\frac{\pd p^A}{\pd u}+ 2a^2\frac{s}{u+r}p^A\simeq \frac{q \gamma^{AB}}{r^2} \p_u \cA_B,
\end{align}
where we shifted the scale factor, since we want to solve by integration for $N_A=\int_\R \p_u \cA_A \pd u$. Next, we observe that we can write the left-hand side as $\frac{\pd}{\pd u}\left(a^2 p^A \right)$ by using $\frac{1}{a}\frac{\pd a}{\pd u}=\frac{s}{u+r}$ and find therefore
\begin{align}
\frac{\pd}{\pd u}   \left(a^2 p^A\right)\simeq \frac{q \gamma^{AB}}{r^2} \p_u \cA_B. \label{Equation}
\end{align}
Introducing $P^A=a^2 p^A$  and assuming that the position of the particle on the large sphere is sufficiently fixed in the interval $u\in(u_1,u_2)$ to pull $\frac{\gamma^{AB}}{r^2}$ out of the $u$-integral, we obtain
\begin{align}
    \Delta P^A\simeq \frac{q\gamma^{AB}}{r^2}\left(\cA_B(u_2,\bth_2)- \cA_B(u_1, \bth_1)\right).
\end{align}
In order to obtain $N_B(\bth)=\cA_B^+(\bth)-\cA_B^-(\bth)$ on the right-hand side, we need, in principle, to take the limits $u_1\to -\infty$ and $u_2\to \infty$ while keeping $\bth_1\simeq \bth_2$:
\begin{align}
\Delta P^A\simeq \frac{q}{r^2}\gamma^{AB}N_B. \label{momentumchangeFLRW}
\end{align}
The limits of infinitely early and late retarded times lead, however, to problems in both the expanding spacetime and in Minkowski space. In the latter case, for a particle initially at rest, any non-vanishing change in momentum $\Delta P^A=\Delta p^A\neq 0$ would lead to an infinitely large change in position due to the infinitely long time interval in which the momentum is non-vanishing. Thus, the approximation of a nearly fixed position would break down and equation \eqref{momentumchangeFLRW} would no longer hold. In the expanding spacetimes, a particle which is initially at rest will also rest at infinitely late retarded times $u\to \infty$ due to the Hubble friction term. To measure a non-vanishing change in momentum, one should, therefore, make the second measurement at finite retarded time $u$. In FLRW spacetimes, there is yet another problem. While the test particle is located at very large $r$, but not at $\scri$, the limit $u\to -\infty$ is only well-defined on $\scri$; at large but finite $r$, one will rather run into the initial singularity. Due to the condition that the radiation dies off at very early and late retarded times \eqref{magneticabsenceenergyfinite} which means that $\mathcal{F}_{uA}=\p_u \cA_A=0$, there will be retarded times $u_i, u_f$  such that $\forall u<u_i: \cA_B(u)=\cA_B^-$ and $\forall u>u_f:\cA_B(u)=\cA_B^+$. It is then enough for the integration to take $u_1<u_i$ and $u_2>u_f$. This makes the above set-up, at least in principle, meaningful. \\
To underline another difference to flat spacetime, it should be emphasized that the change in angular velocity would, up to a factor $\frac{1}{m}$, be $\Delta p^A$. Here, we obtain a difference $\Delta P^A=a^2(u_2)p^A(u_2)-a^2(u_1)p^A(u_1)$ which is, in comparison to $\Delta p_A$, further affected by the expansion of the universe. \\
Next, we would like to see how the memory effect is sourced. To this end, consider the part of Maxwell's equations that is normal to the constant $v$-hypersurface $\scri$ by evaluating $n^\nu \nabla^\mu F_{\mu \nu}=n^\nu J_\nu$ with the normal vector $n^\mu=\delta^\mu_u-\frac{1}{2}\delta^\mu_r$. Imposing the radial fall-off conditions  \eqref{FLRWcurrentfalloff} and \eqref{fieldstrengthfalloffsFLRW}, we find to leading order
\begin{align}
    \p_u \mathcal{F}_{ru} +\gamma^{AB} D_A \mathcal{F}_{uB}=\lim_{r\to\infty} a^2 r^2\left(J_u-\frac{1}{2}J_r\right)=\mathcal{J}_u.
\end{align}
Imposing also the fall-offs for the gauge fields in the retarded radial gauge \eqref{EDfalloffconditionsFLRW} and integrating over all $u$, we find
\begin{align}
    \gamma^{AB} D_A N_B(\hat{\bx})=\mathcal{F}_{ur}^+(\hat{\bx})-\mathcal{F}_{ur}^-(\hat{\bx})+\int_\R \mathcal{J}_u(u,\hat{\bx})\pd u.
\end{align}
Due to the absence of radial magnetic fields at null infinity $\mathcal{F}_{AB}=0$, we can write $N_B=\p_B N$. As in flat spacetime \cite{Bieri}, we split this memory vector into an ordinary part and a null part according to
\begin{subequations} \label{ambiguousmemory}
\begin{align}
    D^2N_\text{ord}(\hat{\bx})=\mathcal{F}_{ur}^+(\hat{\bx})-\mathcal{F}_{ur}^-(\hat{\bx})-\Braket{\mathcal{F}_{ur}^+(\hat{\bx})-\mathcal{F}_{ur}^-(\hat{\bx})}_{S^2},\label{Nord} \\
    D^2 N_\text{null}(\hat{\bx})=\int_\R \mathcal{J}(u,\hat{\bx})\pd u-\Braket{\int_\R \mathcal{J}(u,\hat{\bx})\pd u}_{S^2} \label{NNull}
\end{align}
\end{subequations}
and such that $N=N_\text{ord}+N_\text{null}$. Since the equation $D^2 \mathcal{G}(\hat{\bx},\hat{\by})=\delta_{S^2}(\hat{\bx}, \hat{\by})-\frac{1}{4\pi}$ can be solved by means of the Green's function $\mathcal{G}(\hat{\bx}, \hat{\by})=\frac{1}{4\pi}\ln(1-\hat{\bx}\hat{\by})$, the memory vectors are found to be
\begin{subequations}
\begin{align}
    N_A^\text{ord}(\hat{\bx})&=-\frac{1}{4\pi} \oint_{S^2} \frac{\left(\mathcal{F}_{ur}^+(\hat{\by})-\mathcal{F}_{ur}^-(\hat{\by})\right)\hat{\by}_A}{1-\hat{\bx}\hat{\by}}\pd \Gamma_y, \label{ordinarymemory}\\
    N_A^\text{null}(\hat{\bx})&=-\frac{1}{4\pi}\oint_{S^2}\int_\R \frac{\mathcal{J}_u(u,\hat{\by}) \hat{\by}_A}{1-\hat{\bx}\hat{\by}} \pd u \pd \Gamma_y, \label{nullmemory}
\end{align}
\end{subequations}
where $\by_A=(\p_A \hat{\bx})\cdot \by$. The homogeneous solutions $N_\text{hom}$ to \eqref{ambiguousmemory} satisfy $D^2 N_\text{hom}=0$, being therefore constants on the sphere and not contributing to the memory vector, which is hence unambiguous. The results formally resemble those in Minkowski space. In the following, we investigate specific examples for both kinds of memory effects. 

\subsubsection*{Ordinary memory effect}
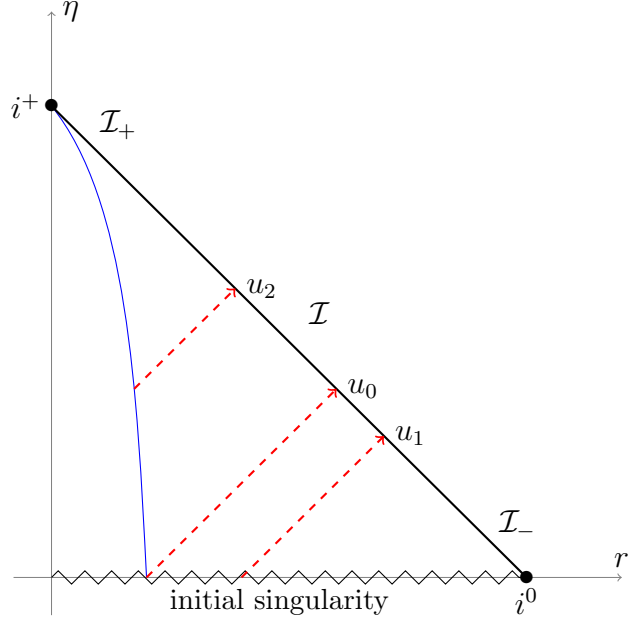
\begin{figure}[h]
		\begin{center}
				\begin{tikzpicture}[scale=2.5]
					\draw[help lines,->] (-0.2,0) -- (3,0) coordinate (raxis);
					\draw[help lines,->] (0,-0.2) -- (0,3) coordinate (taxis);
					\draw[blue] (0.5,0) .. controls (0.45,1) and (0.4,2) .. (0,2.5);
					\draw[thick] (2.5,0) -- (0,2.5);
					\draw[snake] (0,0) -- (2.5,0);
					\draw[thick, dashed, red, ->] (0.5,0) -- (1.5,1);
					\draw[thick, dashed, red, ->] (1,0) -- (1.75,0.75);
					\draw[thick, dashed, red, ->] (0.435,1) -- (0.9675,1.5325);
					
					\node[above] at (raxis) {$r$};
					\node[right] at (taxis) {$\eta$};
					\node[below] at (1.2,0) {{\small initial singularity}};
					\node at (0,2.5) {$\bullet$};
					\node[left] at (0,2.5) {$i^+$};
					\node[right] at (0.2,2.4) {$\mathcal{I}_+$};
					\node at (2.5,0) {$\bullet$};
					\node[below] at (2.5,0) {$i^0$};
					\node[right] at (1.5,1){$u_0$};
					\node[right] at (1.75,0.75){$u_1$};
					\node[right] at (0.9675, 1.5325){$u_2$};
					\node[right] at (2.3,0.28) {$\mathcal{I}_-$};
					\node[right] at (1.3,1.4) {$\mathcal{I}$};
				\end{tikzpicture}
		\end{center}
		\caption{A massive electrically charged particle moves along a timelike worldline (blue) and sources a radial electric field $\mathcal{F}_{ur}$ at $\scri$. For times $u>u_0$, e.g. $u_2$, one can find the source of $\mathcal{F}_{ur}$ by following a null geodesic (red, dashed line) from $\scri$ backwards. For $u<u_0$, e.g. $u_1$, this is not possible, in contrast to flat spacetimes.}
		\label{fig:diagram2}
	\end{figure}

In the case of the ordinary memory effect, the transition from initial to final radiative vacuum is sourced by a change of the radial electric field at $\scri$ from $\mathcal{F}_{ur}^-$ to $\mathcal{F}_{ur}^+$ as can be seen from \eqref{ordinarymemory}. In flat spacetime, such a variation can be caused by a change in the motion of a massive charged point particle \footnote{This point particle is not the test particle at large distances.}. The field strength tensor associated with such a point particle is given by the well-known Liénard-Wiechert solutions. For FLRW spacetimes, closely related solutions are derived in appendix \ref{LienardWiechertLikeSolutionsforFLRW}. According to this, a particle of charge $Q$ moving with (time-dependent, rescaled) peculiar velocity $\bv_Q=a\frac{\pd \bx_Q}{\pd t_Q} $ and spacetime trajectory given by $(\eta, \bx)=(\eta_Q, \bx_Q)$, creates a radial electric field of the form (cf. \eqref{FRI})
\begin{small}
\begin{align}
    F_{\eta i}(x)=\frac{Q}{4\pi(R-\bR\bv_Q)^2}\left\{ \frac{1}{\gamma^2} \frac{\bR^i-R\bv_Q^i}{R-\bR\bv_Q}+a(\eta_Q)\left(\frac{\bR\dot{\bv}_Q}{R-\bR\bv_Q}(\bR^i-R\bv_Q^i)-R\dot{\bv}_Q^i\right)\right\}\bigg\vert_{\eta_Q=\eta-R(\eta_Q)} \label{LWSolution}.
\end{align}
\end{small}
Notice that the dependence on the point where the field strength is evaluated is contained in $\bR(\eta_Q)=\bx-\bx_Q(\eta_Q)$. Moreover, a dot denotes a derivative with respect to time $t_Q$ as measured by an observer comoving with the cosmic fluid.\\
In flat spacetime it is reasonable to assume the particle's velocity to be initially and finally constant. There, this means that the particle moves in the beginning and in the end along a geodesic. For FLRW spacetimes, the velocity is not constant on a geodesic, with exception of vanishing velocity. The spatial part of the geodesic equation implies for the peculiar velocity of a particle with velocity $\bv_0$ at $t_0$
\begin{align}
\bv(t)= \frac{a(t_0)}{a(t)} \frac{\bv_0}{\sqrt{1-\left(1-\frac{a^2(t_0)}{a^2(t)}\right)v_0^2}}.
\end{align}
For the considered decelerating FLRW spacetimes, we have $a(t)=\left(\frac{t}{t_0}\right)^h$ with $0<h<1$. Hence, we find from the previous expression for the velocity that a particle moving along a geodesic always approaches a vanishing velocity at very late times $\lim_{t\to\infty} \bv(t)=\boldsymbol{0}$. This is independent of its initial velocity $\bv_0$. \\
To obtain the leading order component $\mathcal{F}_{ur}$ from \eqref{LWSolution}, we multiply it with the radial outward pointing unit vector $\hat{\bx}$ and take the limit to $\scri$. Since $\bx_Q$ will stay for all times infinitely far away from $\scri$, this means $\bR\to \bx=r\hat{\bx}$ and hence $R\to r$. Thereby, we get for the retarded time in this limit $\eta_Q=\eta-R(\eta_Q)\to \eta-r=u$:
\begin{align}
    \mathcal{F}_{u r}(x)=\frac{Q}{4\pi}\frac{1-\bv_Q^2(u)}{ (1-\hat{\bx}\bv_Q(u))^2},
\end{align}
where $\bv_Q(u)$ means that we replaced $\eta_Q$ by $u$ which is therefore different from $\bv_Q(\eta_Q(u))$. Note that this solution is the same we would have obtained if we had assumed that $\frac{\pd \bv_Q}{\pd t}=0$. We presume that the particle will move along a geodesic at late times. Accordingly, we know from our previous discussion that the particle will eventually come to rest  $\lim_{u\to \infty}\bv_Q(u)=\boldsymbol{0}$ and, therefore,
\begin{align}
    \mathcal{F}_{ur}^+(x)=\frac{Q}{4\pi}.
\end{align}
If we now started with an initially resting particle, there would be no memory effect because the source term in equation \eqref{ordinarymemory} would vanish. To arrive at an interesting result, it is preferable to assume a particle which ``enters" at retarded time $u_a$ with velocity $\bv$. For the relation to $N_B$, we need then also to assume that $\mathcal{F}_{ur}\big\vert_{\scri_-}=\mathcal{F}_{ur}(u_a)$. In this case, we would end up with the equation
\begin{align}
    D^2 N_\text{ord}(\hat{\bx})=\frac{Q}{4\pi}\left(1-\frac{1-\bv^2}{(1-\hat{\bx}\bv)^2}\right) ,
\end{align}
which is solved by
\begin{align}
  N_A=\frac{Q}{4\pi} \frac{\bv_A}{1-\bv\hat{\bx}}.
\end{align}
Here, $\bv_A=(\p_A\hat{\bx})\cdot \bv$ is $r$-independent. We eventually obtain
\begin{align}
    a^2(\eta(u_2))p^A(u_2)-a^2(\eta(u_1))p^A(u_1)\simeq \frac{qQ \gamma^{AB}}{4\pi r^2} \frac{\bv_B}{1-\bv\hat{\bx}}. \label{ofthe} 
\end{align}
The ordinary memory effect exists, therefore, in a similar form to the one in flat spacetime. It should be emphasized once again at this point that the ``memory quantity" is no longer $p^A$ itself but rather $P^A=a^2 p^A$. In addition to the appearance of the scale factor on the left side of \eqref{ofthe}, there is another striking difference. To observe this, recall that the charge $Q$, which produces the radiation by acceleration, is located at much smaller $r$ than the test particle. For the most dramatic case, let it initially rest at $r=0$. Then, along the worldline of the particle, we have initially $\eta=u$. In the limit $\eta\to 0$, and so $u\to 0$, we approach the initial singularity. Hence, for times $u<0$, it is not meaningful to consider the radial electric field at null infinity $\mathcal{F}_{ur}$, which is generated by the charge $Q$. This situation is illustrated in figure \ref{fig:diagram2} and makes the setup slightly more artificial than in flat spacetime.

\subsubsection*{Null memory effect}
In case of the null memory effect, it is convenient to start from the general solution \eqref{nullmemory}.
As a concrete example, consider a massless point particle of charge $Q$ with constant four-velocity $u^\mu=(1,\hbc)$ moving along a worldline with constant retarded time $u=0$. Hence, its four-current density is of the form $J^\eta=\frac{Q}{a^4}\delta^{(3)}(\bx-\hbc\eta)$, $J^i=\frac{Q}{a^4}\hbc^i \delta^{(3)}(\bx-\hbc \eta)$. The particle exits the large sphere at the location specified by $\hbc$. In retarded Bondi coordinates, one finds
\begin{align}
    J_u(u,r,\hat{\bx})=\frac{Q}{a^2 r^2}\delta(u)\delta_{S^2}(\hat{\bx}, \hbc), && J_r=0 .
\end{align}
Plugging this source into \eqref{nullmemory} yields 
\begin{align}
    N_A^\text{null}(\hat{\bx})=-\frac{Q}{4\pi} \frac{\hbc_A}{1-\hat{\bx}\hbc},
\end{align}
which is, as expected from a massless particle, of the same form as in Minkowski space.
According to \eqref{momentumchangeFLRW}, the kick-memory effect is still a change in $a^2 p^A$ and not in $p^A$ as it would have been in flat spacetime. This is because the charged test particle is massive and, hence, breaks the conformal invariance. The measurable change is consequently
\begin{align}
    a^2(\eta(u_2))p^A(u_2)-a^2(\eta(u_1))p^A(u_1)=\frac{qQ\gamma^{AB}}{4\pi r^2}.
\end{align}

\subsubsection*{Remark: Displacement memory effect}
In the case of flat spacetime, in addition to the kick-memories, a modified setting was introduced, e.g. in \cite{Pasterski, Miller}, which we will briefly review here. In this setup, the test particle is immersed in a viscous fluid to keep it close to its original position and measure the memory as a displacement rather than a kick. Let us assume that the friction term is, as suggested by Stokes' law, linear in the particles momentum. Using the same approximations as before, we find from the Lorentz force equation 
\begin{align}
    \frac{\pd p^A}{\pd u}\simeq \frac{q\gamma^{AB}}{r^2}\p_u \cA_B -f p^A, \label{kickmemoryeq2}
\end{align}
where $f>0$ is a constant, describing the friction strength. Because there is no radiation at $\scri^+$ at very late or early retarded times, it follows that in both these time regimes the momentum will be damped as $\sim\exp(-fu)$, so $\lim_{u\to\pm \infty} p^A(u)=0$ and, thus, $\Delta p^A\simeq 0$. An integration of the previous equation then leads to a displacement on the sphere \cite{Miller}
\begin{align} 
    \Delta x^A= \frac{q\gamma^{AB}}{fmr^2}N_B, \label{deltax}
\end{align}
even if we take $u_i\to-\infty$, $u_f\to \infty$.\\
Unfortunately, this does not work in the case of an expanding universe. If we introduced a friction term proportional to the test-particle's momentum, our approximations would lead us to an equation of the form
\begin{align}
    \frac{\pd}{\pd u}\left(a^2p^A\right)\simeq \frac{q\gamma^{AB}}{r^2}\p_u \cA_B - f a^2 p^A.
\end{align}
To obtain the memory vector $N_A$, we would integrate this equation over an interval $(u_1, u_2)$ and then take this interval large enough to reach the initial and final configuration of vanishing radiation. The left-hand side would then, as in flat spacetime, decay exponentially with $u_1$ and $u_2$, respectively, so that we could neglect it. On the right-hand side, however, we would not simply obtain a shift in the position of the test particle this time, but rather the integral $\sim\int_{u_1}^{u_2} a^2 p^A \pd u$. Unfortunately, this does not imply a substantial improvement and hence does not seem suitable for the cosmological case.

\section{Yang-Mills theory}  
\label{Sec:YM}

In this section, we begin by considering asymptotic symmetries of non-Abelian gauge theories and briefly comment on the associated memory effects. Right after, we discuss asymptotic symmetries in a novel setting where $SU(\mathcal{N})$-Yang-Mills is dynamically coupled to a massless Dirac field and a massless conformally coupled complex scalar field \footnote{It should be noted that we are studying a classical and not a quantum field theory. In this context, the following considerations should only be seen as a first step to explore the asymptotic structure because purely classical Yang-Mills theories (except from electrodynamics) have no physical relevance at present, mainly because they neglect the running of the coupling constant. Nevertheless, in order to understand formal aspects of the theory and as a component of the infrared triangle, these investigations have an undeniable relevance. Besides, such studies can shed light on the connection between gauge theories and gravity, such as the classical double copy \cite{Monteiro:2014cda,Campiglia:2021srh, Adamo:2021dfg}.}. 

\subsection{Asymptotic symmetries}
\label{Subsec:YangMillsSymmetries}
We begin with the study of asymptotic symmetries of Yang-Mills theories. The discussion adds detail to \cite{HeMitraCov}, although the chosen boundary conditions at $\scri_+$ and the obtained Poisson brackets deviate.\\
We consider the gauge group $SU(\mathcal{N})$, such that the gauge field $A_\mu$, usually called gluon field, takes values in the associated Lie algebra $\mathfrak{su}(\mathcal{N})$ with Lie brackets $\llbracket\cdot, \cdot\rrbracket$. We denote a set of generators of the algebra by $T^\fa\in \mathfrak{su}(\mathcal{N})$ which satisfy 
$\llbracket T^\fa, T^\fb\rrbracket =i\sum_{\fc=1}^{\mathcal{N}^2-1} f^{\fa\fb\fc} T^\fc\equiv if^{\fa\fb\fc} T^\fc$. The Lie algebra generators are normalized such that $\tr(T^\fa T^\fb)=\frac{1}{2}\delta^{\fa\fb}$.  Furthermore, the field strength is defined as $F_{\mu\nu}\equiv 2\p_{[\mu} A_{\nu]}-ic \llbracket A_\mu, A_\nu\rrbracket$ or, equivalently, $F_{\mu\nu}^\fa=2\p_{[\mu} A_{\nu]}^\fa +c f^{\fa\fb\fc}A_\mu^{\fb} A_\nu^{\fc} $ where $c$ denotes the coupling constant. The action of Yang-Mills theory can then be written as
\begin{align}
S_\text{YM}=-\frac{1}{2}\int_\mathcal{M} \pd^4 x \sqrt{-g} \tr\left(F^{ \mu\nu} F_{\mu\nu}\right).\label{YMaction}
\end{align}
It is conformally invariant and also invariant under gauge transformations of the form
\begin{align}
A_\mu(x)\longrightarrow \Lambda(x) A_\mu(x) \Lambda^{-1}(x) + \frac{i}{c}\Lambda(x) \p_\mu \Lambda^{-1}(x) \label{YMgaugetrafo}
\end{align}
where $\forall x\in \mathcal{M}: \Lambda(x)\in SU(\mathcal{N})$. We can rewrite this group element via the exponential map as $\Lambda(x)=\exp\left(ic\lambda(x)\right)$ where $\lambda(x)\in \mathfrak{su}(\mathcal{N})$ is the gauge parameter. The equations of motion read
\begin{align}
\widehat{\nabla}^\mu F_{\mu\nu}=ic \llbracket A^\mu, F_{\mu\nu}\rrbracket . \label{YangMillsequations}
\end{align}

We choose the retarded radial gauge $A_r=0$ and $A_u\big\vert_{\scri}=0$. The residual gauge parameter is indeed independent of $r$ and on $\scri$ also of $u$. This can be seen from $0=\delta_\epsilon A_r=\p_r \epsilon+ic\llbracket \epsilon, A_r\rrbracket=\p_r \epsilon$ and in the same way for $A_u$ evaluated on $\scri$. Due to the conformal invariance of the theory, we can apply the same reasoning we used for the scalar field in the paragraph below \eqref{largerconditionscalarfield} to argue that our second finiteness condition takes the form $\left|\int_\scri T_{\mu\nu}(F) X^\mu \pd \Sigma^\nu\right|=\left|\int_{\scri^+} T_{\mu\nu}^\mathbb{M}(F) X^\mu \pd \Sigma^\nu_{\mathbb{M}}\right|<\infty$. As in the case of electrodynamics, together with the equations of motion, this imposes the constraints \cite{Pate,HeMitraCov}
\begin{align}
    F_{ur}\in \sO(r^{-2}), && F_{uA}\in \sO(r^0), && F_{Ar}\in \sO(r^{-2}), && F_{AB}\in\sO(r^0). \label{YMradialfalloffs1}
\end{align}
We will further assume that vacuum configurations are approached at times $u<u_i$ and $u>u_f$, i.e., there $\mathcal{F}_{uA}=\mathcal{F}_{AB}=0$ \cite{Pate}.\\
We choose the naturally suggested radial fall-off conditions for the gauge fields in retarded radial gauge as
\begin{align}
    A_u\in \sO(r^{-1}), && A_r=0, && A_B\in \sO(r^0). \label{YMfalloffs2}
\end{align}
The above conditions on the field strength imply for the gauge fields to leading order $\cA_B(u<u_i, \bth)=\cA_B^-(\bth)$, $\cA_B(u>u_f, \bth)=\cA_B^+(\bth)$ and, furthermore, the existence of two Lie algebra valued functions $\alpha^\pm$, such that $\cA_B^\pm(\bth)=\frac{i}{c}e^{ic\alpha^\pm(\bth)}\p_B e^{-ic\alpha\pm(\bth)}$. This means, the field is pure gauge there.\\ 
From the boundary term of the variation of the action \eqref{freeEDactionFLRW}, we can read off the symplectic potential, and hence the symplectic current, as
\begin{align}
    \Theta_\mu=- F_{\mu\nu}^\fa \delta A^{\fa\nu}, \\
    \omega_\mu=-\left(\delta(\p_\mu A_\nu^\fa)-\delta(\p_\nu A_\mu^\fa)+c f^{\fa\fb\fc}\left(\delta A_\mu^\fb A_\nu^\fc+A_\mu^\fb \delta A_\nu^\fc\right)\right)\curlywedge \delta A^{\nu\fa} , \label{YMsymplecticcurrent}
\end{align}
by means of which we obtain the pre-symplectic form
\begin{align}
    \tilde{\Omega}_{\scri}=\int_{\scri} \omega_\mu \pd \Sigma^\mu=\int_{\scri} \gamma^{AB} \delta(\p_u \cA_A^\fa)\curlywedge \delta \cA^a_B \pd u\wedge \pd\Gamma,
\end{align}
where we used the radial fall-off conditions \eqref{YMradialfalloffs1}, \eqref{YMfalloffs2}. Because of these, the explicitly non-Abelian term in \eqref{YMsymplecticcurrent} drops out of the pre-symplectic form. As in electrodynamics, this pre-symplectic form is weakly degenerate. To resolve the issue, one can extend the phase space in a similar manner as in section \ref{Subsec:asympsym}. We first separate out the $u$-independent part of the gauge field $G_B^\fa(\bth)=\frac{1}{2}(\cA^{+\fa}_B(\bth)+\cA^{-\fa}_B(\bth))$ and write $\cA^\fa_B(u,\bth)=\hat{\cA}_B^\fa (u,\bth)+G^\fa_B(\bth)$. We define the soft gluon field via $N_B^\fa(\bth)=\int_{\R} \p_u \cA_B^\fa(u,\bth)\pd u$. The new fields inherit their transformation behavior under asymptotic symmetries from the transformation behavior of the full gauge field. By construction, $G_B$ transforms as $\cA_B$, while $\hat{\cA}_B$ and $N_B$ transform like the field strength. Under infinitesimal symmetries the transformations are as follows
\begin{subequations}\label{seveneleven}
\begin{align}
    \cA_B^\fa&\longrightarrow \cA_B^\fa+\p_B \epsilon^\fa +cf^{\fa\fb\fc} \cA_B^\fb \epsilon^\fc, \\
    G_B^\fa&\longrightarrow G_B^\fa +\p_B \epsilon^\fa +cf^{\fa\fb\fc} G_B^\fb \epsilon^\fc, \\
    \hat{\cA}^\fa_B&\longrightarrow \hat{\cA}_B^\fa +cf^{\fa\fb\fc} \hat{\cA}^b_B \epsilon^\fc, \\
    N^\fa_B &\longrightarrow N^{\fa}_B +cf^{\fa\fb\fc}N_B^\fb \epsilon^\fc.
\end{align}
\label{YangMillstransformationbehaviour}
\end{subequations}
The symplectic form on the extended phase space parametrized by $\hat{\cA}_B, G_B, N_B$ is then
\begin{align}
    \Omega_{\scri} &=\int_{\scri} \gamma^{AB} \delta^{\fa\fb} \delta(\p_u \hat{\cA}_A^\fa(u,\bth))\curlywedge \delta \hat{\cA}_B^\fb(u,\bth) \pd u \wedge \pd \Gamma +\nonumber\\&+\oint_{S^2} \gamma^{AB}\delta^{\fa\fb} \delta N_A^\fa(\bth)\curlywedge \delta G_B^\fb(\bth) \pd \Gamma.
\end{align}
In the exact same way as for electrodynamics, we can now find the Poisson brackets which are
\begin{align}
    \Set{\hat{\cA}^\fa_A(u,\bth), \hat{\cA}_B^\fb(u',\bth')}&=-\frac{1}{4}\gamma_{AB}\delta^{\fa\fb} \sgn(u-u')\delta_{S^2}(\bth, \bth'), \\
    \Set{G_A^\fa(\bth), N_B^\fb(\bth')}&= \gamma_{AB}\delta^{\fa\fb} \delta_{S^2}(\bth, \bth').
\end{align}
While these brackets are fine on their own, it must be noted that they exhibit the same inconsistency described for electrodynamics if one wants to express $N_B$ again in terms of $\cA_B$ (cf. discussion below \eqref{U1chargealgebra}).\\
Our next step is to derive the asymptotic charges. This can be either done under application of Noether's theorem or via the procedure laid out at the end of section \ref{Sec:CPS}, the latter being followed here. The phase space vector field generating the asymptotic symmetries is obtained from the transformation behaviors \eqref{YangMillstransformationbehaviour} as 
\begin{align}
    \Upsilon_\epsilon&=\int_{\scri} cf^{\fa\fb\fc} \hat{\cA}_B^\fb(u,\bth) \epsilon^\fc(\bth) \frac{\delta}{\delta \hat{\cA}^\fa_B(u, \bth)}\pd u\wedge \pd \Gamma +\oint_{S^2} c f^{\fa\fb\fc} N^\fb_B(\bth)\epsilon^\fc(\bth) \frac{\delta}{\delta N_B^\fa(\bth)}\pd \Gamma +\nonumber\\&+\oint_{S^2} \left(\p_B\epsilon^\fa(\bth) + cf^{\fa\fb\fc}G_B^\fb(\bth)\epsilon^\fc(\bth)\right) \frac{\delta}{\delta G^\fa_B(\bth)}\pd \Gamma.
\end{align}
The asymptotic charge can then be found from \eqref{secondchargedefinition} as 
\begin{align}
    \mathcal{Q}[\epsilon]&=-cf^{\fa\fb\fc}\int_{\scri} \gamma^{AB} \hat{\cA}_A^\fa(u,\bth) \p_u \hat{\cA}_B^\fb(u,\bth) \epsilon^\fc(\bth) \pd u \wedge \pd \Gamma-\nonumber \\ &- cf^{\fa\fb\fc}\oint_{S^2} \gamma^{AB} G_A^\fa(\bth) N_B^\fb(\bth) \epsilon^\fc(\bth) \pd u \wedge \pd \Gamma+
    \oint_{S^2} \gamma^{AB} N_A^\fa(\bth) \p_B \epsilon^\fa(\bth) \pd \Gamma \equiv \nonumber\\&\equiv \mathcal{Q}^\text{NA}_\text{H}[\epsilon]+\mathcal{Q}^\text{NA}_\text{S}[\epsilon]+\mathcal{Q}^\text{A}_\text{S}[\epsilon], \label{asymptoticchargeYMMink}
\end{align}
where we denote the terms, respectively, as non-Abelian hard charge, non-Abelian soft charge and Abelian soft charge. Once again, the name is due to the fact that the soft charges are only expressed by fields whose energy vanishes, while the energy of the fields in the hard charge does not vanish. We consistently obtain the Abelian case by setting the structure constants equal to zero. The reason that we then do not have a hard charge anymore lies in the fact that, unlike in our treatment of electrodynamics, we have not coupled the gauge field to a charged current.\\
By construction, the charges generate the asymptotic symmetries
\begin{subequations}
\begin{align}
    \Set{\mathcal{Q}[\epsilon], N_A^\fa(\bth)}=\Set{\mathcal{Q}_\text{S}^\text{NA}[\epsilon], N_A^\fa(\bth)}&=-\delta_\epsilon N_A^\fa(\bth), \\
    \Set{\mathcal{Q}[\epsilon], G^\fa_A(\bth)}=\underbrace{\Set{\mathcal{Q}_\text{S}^\text{NA}[\epsilon], G_A^\fa(\bth)}}_{=-cf^{\fa\fb\fc} G_B^\fb(\bth) \epsilon^\fc(\bth)}+\underbrace{\Set{\mathcal{Q}_\text{S}^\text{A}[\epsilon], G_A^\fa(\bth)}}_{=-\p_A \epsilon^\fa(\bth)}&=-\delta_\epsilon G_A^\fa(\bth), \\
    \Set{\mathcal{Q}[\epsilon], \hat{\cA}_A^\fa(u,\bth)}=\Set{\mathcal{Q}^\text{NA}_\text{H}[\epsilon], \hat{\cA}^\fa_A(u,\bth)}&=-\delta_\epsilon \hat{\cA}_A^\fa (u,\bth).
\end{align}
\end{subequations}
Using that $\llbracket\cdot, \cdot\rrbracket$ are Lie brackets, one can also find that the following charge algebra relations (up to vanishing brackets) hold
\begin{subequations}
\begin{align}
    \Set{\mathcal{Q}_\text{H}^\text{NA}[\epsilon],\mathcal{Q}_\text{H}^\text{NA}[\epsilon']}&=ic\mathcal{Q}_\text{H}^\text{NA} [\llbracket\epsilon, \epsilon'\rrbracket], \\
    \Set{\mathcal{Q}^\text{NA}_\text{S}[\epsilon], \mathcal{Q}^\text{NA}_\text{S}[\epsilon']}&=ic\mathcal{Q}^\text{NA}_\text{S}[\llbracket\epsilon, \epsilon'\rrbracket],\\
    \Set{\mathcal{Q}^\text{NA}_\text{S}[\epsilon], \mathcal{Q}_\text{S}^\text{A}[\epsilon']}+\Set{\mathcal{Q}_\text{S}^\text{A}[\epsilon], \mathcal{Q}^\text{NA}_\text{S}[\epsilon']}&=ic\mathcal{Q}^\text{A}_\text{S}[\llbracket \epsilon, \epsilon'\rrbracket], \\
    \Set{\mathcal{Q}[\epsilon], \mathcal{Q}[\epsilon']}&=ic\mathcal{Q}[\llbracket \epsilon, \epsilon'\rrbracket].
\end{align}
\end{subequations}

\subsubsection*{Comment on memory effects}
So far, we have only taken a look at memory effects in the context of electrodynamics. Nevertheless, the kick-memory effects are modified in a similar way in Yang-Mills theories. A scalar test particle of color charge $q$ and mass $m$ is governed by the curved spacetime generalization of the Wong equations \cite{Wong}
\begin{align}
 m\frac{\text{D}u^\alpha}{\pd \tau}=2cg^{\alpha\beta}\tr\left(F_{\beta\mu}q\right)u^\mu   && \frac{\pd q}{\pd \tau}=icu^\mu \llbracket A_\mu, q\rrbracket. \label{Wong1equations} 
\end{align}
With the slow-motion and large-$r$ approximations, analogous to the considerations in the case of electrodynamics, these equations take in the first non-vanishing order the form
\begin{align}
    \frac{\pd}{\pd u}\left(a^2 p^A\right)\simeq\frac{2c\gamma^{AB}}{r^2}\tr\left(\mathcal{F}_{uB}q\right), && \frac{\pd q}{\pd u}\simeq\frac{ic}{r} \llbracket \cA_u, q\rrbracket. 
\end{align}
The first of these two equations adapts in the same way as in electrodynamics, meaning that it is modified with respect to flat spacetime by replacing $p^A$ with $a^2p^A$. The second of the above equations remains formally the same as in Minkowski space and the charge stays constant at leading order in the large-$r$ expansion in retarded radial gauge. Since $\mathcal{F}_{uB}=\p_u \cA_B$, the Yang-Mills kick-memory is given by the equation
\begin{align}
    \Delta P^A\simeq\frac{2c\gamma^{AB}}{r^2}\tr\left(qN_B\right).
\end{align}
Due to Yang-Mills equations being conformally invariant, $N_B$ is still determined by the same equation as in Minkowski space
\begin{align}
    \gamma^{AB}D_A \p_u \cA_B=ic \gamma^{AB} \llbracket \cA_A, \p_u \cA_B\rrbracket + \p_u \mathcal{F}_{ur},
\end{align}
upon integration over $u$ (cf. e.g. \cite{Pate}).

Before we continue, let us briefly note that other memory effects like phase memories and color memories have been examined in detail in flat spacetimes, e.g. \cite{Pate, Campoleoni, Jokela}. In order to analyze these effects, it is necessary to transition to a quantum mechanical treatment. To apply standard quantum mechanics as a first approximation to the memories in our curved spacetimes, we need to assume that the experiment happens on time-scales over which the scale factor stays approximately constant. In such a case, the results in FLRW will formally resemble those in Minkowski. A treatment including the quantum effects due to the expansion of the universe is beyond the scope of this manuscript. Nonetheless, it might be a promising starting point for future research.

\subsection{Phase space with fermions}
\label{subsec:fermionicps}
In this subsection, we dynamically couple the $SU(\mathcal{N})$-Yang-Mills field to a massless Dirac field $\bpsi=\left(\Psi_1,\dots, \Psi_{\mathcal{N}}\right)$ and a massless conformally coupled complex scalar field $\bphi=(\Phi_1, \dots, \Phi_{\mathcal{N}})$, both in the fundamental representation. We generally use compact notation, e.g. $\bphi^\dagger \bphi=\sum_{\mathfrak{i}=1}^{\mathcal{N}} \Phi^*_\mathfrak{i} \Phi_\mathfrak{i}$ or
$\bar{\bpsi}\bpsi=\sum_{\mathfrak{i}=1}^{ \mathcal{N}} \Bar{\Psi}_\mathfrak{i} \Psi_\mathfrak{i}$. Lower indices $\mathfrak{i},\mathfrak{j}\in \Set{1,\dots,\mathcal{N}}$ of the matter fields are associated with the fundamental representation.

The action of the theory we consider is given by 
\begin{align}
    S=\int_{\mathcal{M}}\sqrt{-g}\pd^4 x \biggl\{-\frac{1}{2}\tr\left(F_{\mu\nu}F^{\mu\nu}\right)+g^{\mu\nu}\nabla_\mu \bphi^\dagger \nabla_\nu \bphi -ic \left(\nabla^\mu \bphi^\dagger A_\mu\bphi -\bphi^\dagger A_\mu\nabla^\mu \bphi\right) +\nonumber\\+c^2 \bphi^\dagger A_\mu A^\mu \bphi +\frac{R}{6}\bphi^\dagger \bphi +\frac{i}{2}\left(\bar{\bpsi}\gamma^\mu \nabla_\mu \bpsi-(\nabla_\mu \Bar{\bpsi})\gamma^\mu \bpsi\right)+c\bar{\bpsi}\gamma^\mu A_\mu \bpsi\biggr\}
 \label{fullaction}\end{align}
and exhibits an invariance under the gauge transformations
\begin{subequations}
\begin{align}
    A_\mu \longrightarrow \Lambda A_\mu \Lambda^{-1}+\frac{i}{c}\Lambda \p_\mu \Lambda^{-1}, && F_{\mu\nu}\longrightarrow \Lambda F_{\mu\nu} \Lambda^{-1},\label{955a}\\
    \bphi \longrightarrow \Lambda \bphi, \;\;\; \bphi^\dagger \longrightarrow  \bphi^\dagger \Lambda^{-1}, &&
    \bpsi\longrightarrow \Lambda \bpsi,\;\;\;  \bpsi^\dagger \longrightarrow \bpsi^\dagger \Lambda^{-1}, \label{955b}
\end{align}
\end{subequations}
where $\forall x\in \mathcal{M}:\Lambda(x)\in SU(\mathcal{N})$. 
We begin with the equations of motion in retarded Bondi coordinates. To exploit the conformal invariance of the theory, we introduce auxiliary fields $\bchi=a\bphi$ and $\bzeta=a^{3/2}\bpsi$ (cf. sections \ref{Subsubsec:Scalar} and \ref{Subsubsec:Dirac}). Furthermore, we fix the retarded radial gauge $A_r=0$ and $A_u\big\vert_{\scri}=0$. This restricts the residual gauge symmetries to $r$- and $u$-independent gauge transformations $\Lambda(\bth)$. In this gauge, the equations of motion read \footnote{We have omitted those equations that are related by Hermitian conjugation to the written ones.}

\begin{small}
\begin{subequations}
\begin{align}
    \left(\p_u-\p_r\right) F_{ru}^\fa-\frac{2}{r}F_{ru}^\fa-\frac{\gamma^{AB}}{r^2}D_A F_{Bu}^\fa=&ic\llbracket A_u, F_{ru}\rrbracket^\fa-ic\frac{\gamma^{AB}}{r^2}\llbracket A_A, F_{Bu}\rrbracket^\fa +ic\left(\p_u \bchi^\dagger T^\fa \bchi-\bchi^\dagger T^\fa \p_u \bchi\right)\nonumber\\&-c\bar{\bzeta}\frac{\gamma_u}{a} T^\fa \bzeta -c^2 \bchi^\dagger\left(T^\fa A_u +A_u T^\fa\right)\bchi,\\
\p_r F_{ur}^\fa+\frac{2}{r}F_{ur}^\fa-\frac{\gamma^{AB}}{r^2}D_A F_{Br}^\fa=&-ic\frac{\gamma^{AB}}{r^2}\llbracket A_A, F_{Br}\rrbracket^\fa +\nonumber\\&+ic\left(\p_r \bchi^\dagger T^\fa \bchi-\bchi^\dagger T^\fa \p_r \bchi\right)-c\bar{\bzeta}\frac{\gamma_r}{a} T^\fa \bzeta ,\\
\left(\p_u-\p_r\right) F_{rC}^\fa+\p_r F_{uC}^\fa-\frac{\gamma^{AB}}{r^2}D_A F_{BC}^\fa=&ic\llbracket A_u, F_{rC}\rrbracket^\fa-ic\frac{\gamma^{AB}}{r^2}\llbracket A_A, F_{BC}\rrbracket^\fa +ic\left(\p_C \bchi^\dagger T^\fa \bchi-\bchi^\dagger T^\fa \p_C \bchi\right)\nonumber\\&-c\bar{\bzeta}\frac{\gamma_C}{a} T^\fa \bzeta -c^2 \bchi^\dagger\left(T^\fa A_C +A_C T^\fa\right)\bchi,
\end{align}
\end{subequations}
\vspace{-0.9cm}
\begin{align}
\left(\frac{2}{r}\left(\p_u-\p_r\right)+2\p_u \p_r -\p_r^2-\frac{1}{r^2}D^2\right)\bchi=&ic\left(2A_u\p_r+\left(\frac{2}{r}+\p_r\right) A_u-\frac{2\gamma^{AB}}{r^2}A_A D_B-\frac{\gamma^{AB}}{r^2}D_A A_B\right)\bchi\nonumber\\&-c^2 \frac{\gamma^{AB}}{r^2}A_A A_B \bchi,
\end{align}
\vspace{-0.9cm}
\begin{subequations}
\begin{align}
    \p_r\left(r\bzeta_L^+\right)&=\left[\frac{\p_\varphi}{\sin(\theta)}-i\p_\theta-\frac{i}{2\tan(\theta)}-ic\left(\frac{A_\varphi}{\sin(\theta)}-iA_\theta\right)\right]\bzeta_L^-,\\
    \left[2\p_u-\p_r-2icA_u\right]\left(r\bzeta_L^-\right)&=\left[\frac{\p_\varphi}{\sin(\theta)}+i\p_\theta+\frac{i}{2\tan(\theta)}-ic\left(\frac{A_\varphi}{\sin(\theta)}+iA_\theta\right)\right]\bzeta_L^+,\\
    \p_r\left(r\bzeta_R^+\right)&=-\left[\frac{\p_\varphi}{\sin(\theta)}+i\p_\theta+\frac{i}{2\tan(\theta)}-ic\left(\frac{A_\varphi}{\sin(\theta)}+iA_\theta\right)\right]\bzeta_R^-,\\
    \left[2\p_u-\p_r-2icA_u\right]\left(r\bzeta_R^-\right)&=-\left[\frac{\p_\varphi}{\sin(\theta)}-i\p_\theta-\frac{i}{2\tan(\theta)}-ic\left(\frac{A_\varphi}{\sin(\theta)}-iA_\theta\right)\right]\bzeta_R^+.
\end{align}
\end{subequations}
\end{small}
Note that in the above equations $\frac{\gamma_u}{a}, \frac{\gamma_r}{a}$, and $\frac{\gamma_C}{a}$ are independent of the scale factor. In appendix \ref{AppendixExpansion}, we show that the fields allow for an asymptotic expansion which is consistent with the above equations of motion. The asymptotic expansion is associated with the following fall-off conditions (for definitions regarding spinor components, see around equation \eqref{plusminusspinor})
\begin{subequations}\label{finalfaloo}
\begin{align} 
    A_u\in \sO\left(r^{-1}\right), && A_r=0, && A_B\in \sO\left(r^{0}\right), \\
    \bphi\in \sO\left(a^{-1}r^{-1}\right), && \bpsi_{L,R}^{+}\in \sO\left(a^{-\frac{3}{2}}r^{-1}\right),&& \bpsi_{L,R}^-\in \sO\left(a^{-\frac{3}{2}}r^{-2}\right).
\end{align}
\end{subequations}
We also show in this appendix that, by specifying $\cA_B(u,\bth)=\lim_{r\to \infty} A_B(u,r,\bth)$, $\bphi^{(1)}(u,\bth)=\lim_{r\to\infty}\left(a(u,r)r\bphi(u,r,\bth)\right)$, and $\bpsi_{L,R}^{+,(1)}(u,\bth)=\lim_{r\to\infty}\left(a^{3/2}(u,r)r\bpsi_{L,R}^+(u,r,\bth)\right)$ \footnote{This is meant to include the Hermitian conjugates of the scalar and spinor fields.} as the boundary data, it is possible to find from the equations of motion the fields to all orders in the asymptotic expansion, upon fixing their value in each order at a certain $u=u_0$ ($u_0$ might be chosen differently for each field and in each order of the expansion). As in section \ref{Subsubsec:Scalar}, we are not going to consider the soft sector of the matter fields and set $\lim_{u\to \pm\infty} \bphi^{(1)}(u,\bth)=0$.

Bearing this in mind, we aim to construct the phase space at null infinity. 
Because of the combination of fermions and bosons, we expect to run into difficulties. Let us present our attempt in the following and discuss the results in detail afterwards.
Since our theory does not only include bosons but also fermions, we associate with each differential form on phase space \footnote{Or configuration space, or solution space, respectively.} a bidegree $(a,b)\in (\mathbb{Z}, \mathbb{Z}_2)$, where $a$ denotes the degree of the form and $b$ its Grassmann parity. \footnote{For a detailed treatment, we refer to \cite{grassmanngraded}.} A bosonic $0$-form like $\bphi$ has for example bidegree $(0,0)$, a fermionic $1$-form like $\delta \bpsi$ has bidegree $(1,1)$. The following rules will follow naturally when keeping in mind that the exchange of two bosons or a boson with a fermion leads to no sign, while the exchange of two fermions yields a sign. More formally, exchanging the order of $\vartheta_{1,2}$ with bidegrees $(0, b_{1,2})$ gives $\vartheta_1\vartheta_2=(-1)^{b_1b_2}\vartheta_2 \vartheta_1$. The wedge product between forms $\vartheta_1, \vartheta_2$ of respective bidegrees $(a_{1,2}, b_{1,2})$ satisfies
\begin{align}
    \vartheta_1 \curlywedge \vartheta_2= (-1)^{a_1 a_2+b_1 b_2} \vartheta_2\curlywedge \vartheta_1.
\end{align}
This ensures that the wedge product of bosonic 1-forms is antisymmetric and the wedge product of fermionic 1-forms is symmetric, which is exactly what we need in order to extract usual Poisson brackets for bosons and symmetric brackets for fermions from the symplectic form. 
Moreover, note that if we have in addition a form $\vartheta_3$ with bidegree $(-1,b_3)$ (vector field), we get for the interior product
\begin{align}
    i_{\vartheta_3}\left(\vartheta_1\curlywedge\vartheta_2\right) =\left(i_{\vartheta_3}\vartheta_1\right)\curlywedge\vartheta_2+(-1)^{a_1+b_1b_3}\vartheta_1 \curlywedge \left(i_{\vartheta_3}\vartheta_2\right).
\end{align}
Equipped with this, we obtain the symplectic current from the variation of the action as
\begin{small}
\begin{align}
   \omega_\mu=&-\left(\delta(\p_\mu A_\nu^\fa)-\delta(\p_\nu A_\mu^\fa)+c f^{\fa\fb\fc}\left(\delta A_\mu^\fb A_\nu^\fc+A_\mu^\fb \delta A_\nu^\fc\right)\right)\curlywedge \delta A^{\nu\fa}+\delta\left(\nabla_\mu \bphi^\dagger\right)\curlywedge \delta \bphi-\delta \bphi^\dagger \curlywedge\delta(\nabla_\mu \bphi)+\nonumber\\&+i\delta\Bar{\bpsi}\curlywedge\left(\gamma_\mu \delta\bpsi\right)+ic\left(\delta\bphi^\dagger\curlywedge \delta A_\mu\bphi+2\delta \bphi^\dagger\curlywedge\left(A_\mu\delta\bphi\right)+\bphi^\dagger \delta A_\mu\curlywedge \delta \bphi\right).
\end{align}
\end{small}
Accordingly, we can obtain the presymplectic form $\tilde{\Omega}_\scri=\int_\scri \omega_\mu \pd \Sigma^\mu$ under application of the fall-off conditions \eqref{finalfaloo}. In doing so, the recursive problem of the gauge field phase space shows up again. This issue can be cured by the procedure of extending the phase space along the lines of section \ref{Subsec:YangMillsSymmetries}, leading to the symplectic form
\begin{align} \label{symplecticextendedphasespace}
    \Omega_\scri=2&\int_{\scri}\gamma^{AB}\tr\left\{\delta(\p_u \hat{\cA}_A)\curlywedge \delta \hat{\cA}_B\right\}\pd u \wedge \pd \Gamma+2\oint_{S^2} \gamma^{AB} \tr\left(\delta N_A\curlywedge\delta G_B\right)\pd \Gamma+\nonumber\\+&2\int_\scri \delta \left(\p_u \bphi^{(1)\dagger}\right)\curlywedge \delta \bphi^{(1)}\pd u\wedge \pd \Gamma+\nonumber\\+&\int_\scri i\left\{\delta \bpsi_L^{+,(1)\dagger}\curlywedge \delta \bpsi_L^{+,(1)}+\delta \bpsi_R^{+,(1)\dagger}\curlywedge \delta \bpsi_R^{+,(1)}\right\}\pd u \wedge \pd \Gamma.
\end{align}
 Note that the symplectic form is of block diagonal form such that the phase space on $\scri$ factorizes into a scalar, a spinor and a gauge field phase space: $\Gamma=\Gamma^\text{YM}\times \Gamma^\Phi\times \Gamma^\Psi$. It should be emphasized again that this holds only due to the asymptotic fall-offs on $\scri$. More generally, mixing between the different spaces could occur.\footnote{Recall that the symplectic form is not uniquely determined in the covariant phase space construction (cf. discussion around equation \eqref{omegadefuniquenon}), which could also contribute to the appearance of a mixing term.} The extended gauge field phase space further factorizes \footnote{Note that this factorization is caused by the phase space extension.} into a part for the $u$-dependent fields and a part for the $u$-independent boundary fields. One obtains the following brackets
\begin{subequations}\label{fundamentalbrackets}
\begin{align}
\Set{\hat{\cA}^\fa_A(u,\bth), \hat{\cA}^\fb_B(u',\bth')}&=-\frac{1}{4}\delta^{\fa\fb}\gamma_{AB}\sgn(u-u')\delta_{S^2}(\bth, \bth'),\\
\Set{G_A^\fa(\bth), N_B^\fb(\bth')}&=\delta^{\fa\fb}\gamma_{AB}\delta_{S^2}(\bth, \bth'),\\
\Set{\Phi^{(1)*}_{\mathfrak{i}}(u,\bth), \Phi^{(1)}_\mathfrak{j}(u',\bth')}&=-\frac{1}{4}\delta_{\mathfrak{ij}}\sgn(u-u')\delta_{S^2}(\bth, \bth'),\\
\Set{\Psi_{L,\mathfrak{i}}^{+,(1)*}(u,\bth),\Psi_{L,\mathfrak{j}}^{+,(1)}(u', \bth')}&=i\delta_{\mathfrak{ij}}\delta(u-u')\delta_{S^2}(\bth,\bth'), \label{fermionbracketa}\\
\Set{\Psi_{R,\mathfrak{i}}^{+,(1)*}(u,\bth),\Psi_{R,\mathfrak{j}}^{+,(1)}(u', \bth')}&=i\delta_{\mathfrak{ij}}\delta(u-u')\delta_{S^2}(\bth,\bth').\label{fermionbracketb}
\end{align}\label{Poissonsuperbrackets}
\end{subequations}
Following the definition of the brackets in \eqref{definitionPoissonbracket} strictly, one can see that the first three brackets between two bosons are antisymmetric, while the last two between two fermions are symmetric.\\
Our final task is now to construct the asymptotic charges and check whether they generate the asymptotic symmetries. This time, our theory contains no external currents but only dynamical fields. Therefore, at first sight, we should be able to obtain the asymptotic charges by application of \eqref{secondchargedefinition}.
First, we need the transformation behavior of the fields under an infinitesimal asymptotic transformation $\Lambda(\bth)=1+ic\epsilon(\bth)$. This, we obtain from  \eqref{seveneleven} and \eqref{955b} as
\begin{subequations}
\begin{align}
&& \delta_\epsilon \hat{\cA}_B(u,\bth)=ic\llbracket \epsilon(\bth), \hat{\cA}_B(u,\bth)\rrbracket,\\
\delta_\epsilon G_B(\bth)=ic\llbracket \epsilon(\bth), G_B(\bth)\rrbracket  +\p_B \epsilon(\bth), && \delta_\epsilon N_B(\bth)=ic\llbracket \epsilon(\bth), N_B(\bth)\rrbracket,\\
\delta_\epsilon \bphi^{(1)}(u,\bth)=ic\epsilon(\bth)\bphi^{(1)}(u,\bth), &&\delta_\epsilon\bphi^{(1)\dagger}(u,\bth)=-ic\bphi^{(1)\dagger}(u,\bth)\epsilon(\bth),\\
\delta_\epsilon \bpsi_{L,R}^{+,(1)}(u,\bth)=ic\epsilon(\bth)\bpsi^{+,(1)}_{L,R}(u,\bth), && \delta_\epsilon \bpsi_{L,R}^{+,(1)\dagger}(u,\bth)=-ic\bpsi^{+,(1)\dagger}_{L,R}(u,\bth)\epsilon(\bth).
\end{align}
\end{subequations}
The phase space vector field $\Upsilon_\epsilon$ generating the asymptotic symmetries, defined such that for any function $\mathfrak{f}\in \mathfrak{F}(\Gamma)$: $\delta_\epsilon \mathfrak{f}=\delta\mathfrak{f}(\Upsilon_\epsilon)$, turns out to be
\begin{footnotesize}
\begin{align}
    \Upsilon_\epsilon=&\int_\scri cf^{\fa\fb\fc}\hat{\cA}_B^\fb(u,\bth)\epsilon^\fc(\bth)\frac{\delta}{\delta \hat{\cA}_B^\fa(u,\bth)}\pd u \wedge\pd\Gamma+\oint\left(cf^{\fa\fb\fc}G_B^\fb(\bth)\epsilon^\fc(\bth)+\p_B\epsilon^\fa(\bth)\right)\frac{\delta}{\delta G_B^\fa(\bth)}\pd \Gamma+\nonumber\\&+\oint_{S^2}c f^{\fa\fb\fc}N_B^\fb(\bth)\epsilon^\fc(\bth)\frac{\delta}{\delta N_B^\fa(\bth)}\pd \Gamma +ic\int_\scri\epsilon_{\mathfrak{ij}}(\bth) \left(\Phi_{\mathfrak{j}}^{(1)}(u,\bth)\frac{\delta}{\delta \Phi_\mathfrak{i}^{(1)}(u,\bth)}-\Phi_{\mathfrak{i}}^{(1)*}(u,\bth)\frac{\delta}{\delta \Phi_\mathfrak{j}^{(1)*}(u,\bth)}\right)\pd u\wedge\pd \Gamma\nonumber\\&-ic\sum_{P\in\Set{L,R}}\int_\scri\epsilon_{\mathfrak{ij}}(\bth)\left(\Psi_{P,\mathfrak{j}}^{+,(1)}(u,\bth)\frac{\delta}{\delta \Psi_{P,\mathfrak{i}}^{+,(1)}(u,\bth)}-\Psi_{P,\mathfrak{i}}^{+,(1)*}(u,\bth)\frac{\delta}{\delta \Psi_{P,\mathfrak{j}}^{+,(1)*}(u,\bth)}\right)\pd u\wedge \pd \Gamma.
\end{align}
\end{footnotesize}
From \eqref{secondchargedefinition}, we obtain the asymptotic charge
\begin{align}
    \mathcal{Q}[\epsilon]=\mathcal{Q}_\text{H}^\text{{NA}}[\epsilon]+\mathcal{Q}_\text{S}^\text{NA}[\epsilon]+\mathcal{Q}_\text{S}^\text{A}[\epsilon]+\mathcal{Q}_\text{H}^\Phi[\epsilon]+\mathcal{Q}^\Psi[\epsilon],
\end{align}
where the first three charges are of the same form as those in \eqref{asymptoticchargeYMMink} and
\begin{subequations}
\begin{align}
    \mathcal{Q}_\text{H}^{\Phi}[\epsilon]=2ic\int_\scri \p_u \bphi^{(1)\dagger}(u,\bth)\epsilon(\bth)\bphi^{(1)}(u,\bth)\pd u\wedge \pd \Gamma,\\
    \mathcal{Q}_\text{H}^\Psi[\epsilon]=-c\sum_{P\in \Set{L,R}}\int_\scri \bpsi_P^{+,(1)\dagger}(u,\bth)\epsilon(\bth)\bpsi_P^{+,(1)}(u,\bth)\pd u\wedge\pd \Gamma. \label{fermioncharge}
    \end{align}
\end{subequations}
These are the same charges one finds by application of Noether's procedure. Using the explicit definitions of the brackets, it is easy to show that they indeed generate the asymptotic symmetries in the sense described at the end of section \ref{Sec:CPS}. 

\subsubsection*{Does the phase space construction lead to Poisson superbrackets?}


Let us now explore the possibility of our brackets being Poisson superbrackets and then check via their properties if the charges generate the symmetries. The defining properties of Poisson superbrackets for arbitrary $(0,b_{1,2,3})$-forms $\vartheta_{1,2,3}$ are
\begin{small}
\begin{subequations}\label{superbrackets}
\begin{align}
&\Set{\vartheta_1, \vartheta_2}=-(-1)^{b_1 b_2}\Set{\vartheta_2, \vartheta_1}, \label{prop1} \\
&0=(-1)^{b_1b_3}\Set{\vartheta_1,\Set{\vartheta_2, \vartheta_3}}+(-1)^{b_2b_1}\Set{\vartheta_2,\Set{\vartheta_3, \vartheta_1}}+(-1)^{b_3b_2}\Set{\vartheta_3,\Set{\vartheta_1, \vartheta_2}}, \label{prop2} \\
&\Set{\vartheta_1, \vartheta_2\vartheta_3}=\Set{\vartheta_1,\vartheta_2}\vartheta_3+(-1)^{b_1b_2}\vartheta_2\Set{\vartheta_1, \vartheta_3}. \label{prop3}
\end{align}
\end{subequations}
\end{small}
The last of these three properties would allow us to calculate the brackets between the fields and the charges from the fundamental brackets \eqref{fundamentalbrackets}. For all the bosonic fields, this yields the correct result but fails for the fermions by a sign. Clearly, the brackets we constructed are not Poisson superbrackets as we had wished. In the following, we will briefly present what we have identified as sources of this issue and options that might be a starting point to understand it.

First, we remark that the brackets we defined already fail to satisfy property \eqref{prop1}. As an example, applying the definition of the brackets, one finds {$\textstyle \Set{\mathcal{Q}_\text{H}^\Psi[\epsilon], \bpsi_R^{+,(1)}}=\Set{\bpsi_R^{+,(1)},\mathcal{Q}_\text{H}^\Psi[\epsilon]}$}. The Grassmann degree of the charge is $0$ and the one of the fermion is $1$. Therefore, we should have obtained, according to \eqref{prop1}, a sign when exchanging the entries in the previous brackets but we did not.  Moreover, sticking strictly to the definition of the brackets in terms of the symplectic form leads to the asymptotic charge algebra
\begin{subequations}
\begin{align}
    \Set{\mathcal{Q}_\text{H}^\Phi[\epsilon], \mathcal{Q}_\text{H}^\Phi[\epsilon']}&=ic\mathcal{Q}^\Phi_\text{H}[\llbracket\epsilon, \epsilon'\rrbracket],\\
    \Set{\mathcal{Q}_\text{H}^{\Psi}[\epsilon],\mathcal{Q}_\text{H}^\Psi[\epsilon']}&=-ic\mathcal{Q}_\text{H}^\Psi[\llbracket\epsilon, \epsilon'\rrbracket], \\
    \Set{\mathcal{Q}^\text{NA}_\text{S}[\epsilon], \mathcal{Q}^\text{NA}_\text{S}[\epsilon']}&=ic\mathcal{Q}^\text{NA}_\text{S}[\llbracket\epsilon, \epsilon'\rrbracket],\\
    \Set{\mathcal{Q}^\text{NA}_\text{S}[\epsilon], \mathcal{Q}_\text{S}^\text{A}[\epsilon']}+\Set{\mathcal{Q}_\text{S}^\text{A}[\epsilon], \mathcal{Q}^\text{NA}_\text{S}[\epsilon']}&=ic\mathcal{Q}^\text{A}_\text{S}[\llbracket \epsilon, \epsilon'\rrbracket], \\
    \Set{\mathcal{Q}[\epsilon], \mathcal{Q}[\epsilon']}&=ic\mathcal{Q}[\llbracket \epsilon, \epsilon'\rrbracket]-2ic\mathcal{Q}_\text{H}^\Psi[\llbracket\epsilon, \epsilon'\rrbracket]\neq ic\mathcal{Q}[\llbracket \epsilon, \epsilon'\rrbracket]. \label{failtoclose}
\end{align}
\end{subequations}
While we had naively expected a closed algebra for the charges $\mathcal{Q}[\epsilon]$, \eqref{failtoclose} fails to close because of the inclusion of fermions. Note that we would have obtained closed charge algebras if we had treated fermionic and bosonic phase spaces separately.

A first possible solution is to deliberately disregard the derivation of the brackets from the symplectic form and assume the fundamental brackets \eqref{fundamentalbrackets} as a definition. Also by definition, one takes them to be Poisson superbrackets, i.e. to satisfy \eqref{superbrackets}. This would lead to $\set{\bpsi_{R,L}^{+,(1)}, \mathcal{Q}_H^\Psi[\epsilon]}=-\delta_\epsilon \bpsi_{R,L}^{+,(1)}$ (similarly for the Hermitian conjugate) and $ \Set{\mathcal{Q}_\text{H}^{\Psi}[\epsilon],\mathcal{Q}_\text{H}^\Psi[\epsilon']}=-ic\mathcal{Q}_\text{H}^\Psi[\llbracket\epsilon, \epsilon'\rrbracket]$, which both still deviate by a sign from the desired result. To make the charge generate the symmetry on the fermions and obtain the correct charge algebra, we would either have to invert the sign of the fermion-charge \eqref{fermioncharge} or to invert the sign for the fermionic brackets \eqref{fermionbracketa} and \eqref{fermionbracketb}. The charge would then generate the asymptotic symmetries and, furthermore, satisfy $\Set{\mathcal{Q}[\epsilon], \mathcal{Q}[\epsilon']}= ic\mathcal{Q}[\llbracket \epsilon, \epsilon'\rrbracket]$. In addition to this being a strongly arbitrary choice, it would not provide a phase space, as we would not have a symplectic form anymore.

Another possibility would be to redefine the brackets. To see this, let us firstly ignore the spacetime and other additional structures which are not directly related to the phase space or the bigrading. We denote fermions by $f$, bosons by $b$ and their associated Hamiltonian vector fields by $F$ and $B$. Our symplectic form \eqref{symplecticextendedphasespace} is then of the schematic structure $\Omega=\delta f_x\curlywedge \delta f_y +\delta b_x\curlywedge \delta b_y\equiv \Omega_f+\Omega_b$. In particular, there are no terms $\delta f_x\curlywedge \delta b_y$, which in general would not be forbidden. The symplectic form is hence Grassmann even, which ensures that the Hamiltonian vector field $B$ to a boson $i_{B}\Omega=-\delta b$ is bosonic and the Hamiltonian vector field $F$ to a fermion $i_{F}\Omega=-\delta f$ is fermionic. We then have the properties
\begin{subequations}
\begin{align}
    \Omega_f(F_1, F_2)=\Omega_f(F_2, F_1), && \Omega_f(F_1, B_2)=\Omega_f(B_2, F_1), && \Omega_f(B_1, B_2)=-\Omega_f(B_2, B_1), \\
    \Omega_b(F_1, F_2)=\Omega_b(F_2, F_1), && \Omega_b(F_1, B_2)=-\Omega_b(B_2, F_1), && \Omega_b(B_1, B_2)=-\Omega_b(B_2, B_1).
\end{align}
\end{subequations}
While we can use fermions to build bosons, we cannot use bosons to build fermions. Therefore, in the case of our symplectic form, $\Omega_b(F_1, B_2)=0$ and $\Omega_b(F_1, F_2)=0$. If we extracted the brackets via $\Set{f_1, b_2}=-\Omega(F_1, B_2)=-\Omega_f(F_1, B_2)$,  they would fail to satisfy property \eqref{prop1}. One may try to modify the definition of the brackets to $\Set{\vartheta_1, \vartheta_2}=-(-1)^{(1-p_1)p_2}\Omega(X_1, X_2)$. Here, $\vartheta_{1,2}$ are $(0,p_{1,2})$-forms and $X_{1,2}$ are the associated Hamiltonian vector fields. On our phase space, the brackets would then satisfy
\begin{align}
    \Set{f_1,f_2}=\Set{f_2, f_1}, && \Set{f_1, b_2}=-\Set{b_2,f_1}, && \Set{b_1, b_2}=-\Set{b_2, b_1}
\end{align}
and hence property \eqref{prop1} would be satisfied, even though the remaining properties \eqref{prop2} and \eqref{prop3} are not guaranteed to hold. This is only true because in our case $\Omega_b(F_1, B_2)=0$ and $\Omega_b(F_1, F_2)=0$ hold, and we have no Grassmann odd part in the symplectic form. While the procedure will not work in general, it might be worth to pursue this approach for the case at hand.

In general, we believe that these issues could be resolved if one treated the phase spaces for the bosons $\Gamma^\text{YM}\times \Gamma^\Phi$ and for the fermions $\Gamma^\Psi$ separately. For the former, the brackets would immediately be Poisson brackets and everything would straightforwardly work out. Furthermore, even though we have not checked it explicitly, it seems to us that appropriate adaptions in section \ref{Sec:CPS} would also allow to build Poisson superbrackets for the fermions alone. Of course, this splitting is only possible if there is no mixing between fermions and bosons such that the space factorizes, which is fortunately the case.

None of the discussed options seems to completely disentangle the clash between a geometrical (phase space) and an algebraic (super-Poisson structure) derivation when incorporating together both bosons and fermions in this classical context. A successful procedure seems to exist in the field of the quantization of gauge systems \cite{Henneaux:1992ig}, which, however, cannot be easily adapted to the formalism presented here. Nonetheless,
the above discussion might provide a starting point for future investigations.

\section{Summary and conclusions}
\label{Sec:Conclusions}

In this paper, we have investigated the asymptotic symmetry and memory effect corners of the infrared triangle for gauge theories, namely electrodynamics and Yang-Mills theory, in decelerating and spatially flat FLRW universes, thus complementing the gravitational analysis for such backgrounds performed in \cite{Heckelbacher,Rojo,Olivieri}. Such an endeavour granted us the opportunity to enhance the state-of-art with a thorough technical analysis, to a degree of detail absent even in Minkowski background, and with a novel setup where Yang-Mills is coupled to both bosons and fermions, raising the question of how to deal with the corresponding phase space.

\subsubsection*{Summary of results}

Let us summarize our results:

\begin{itemize}
    \item In section \ref{Sec:decFLRW}, we explicitly derived a directed volume form on null infinity which is valid for both Minkowski and FLRW spacetimes, filling an existing gap in the literature. The methodology we followed can be easily extrapolated to generic null boundaries.
    
    \item In section \ref{Subsec:currents}, we studied the radial fall-offs of conformally coupled complex scalar fields and Dirac fields in FLRW backgrounds. These are compatible with their respective flat limits. Nevertheless, our results for the Dirac spinor differ with respect to those of \cite{MitraPhD}. Concretely, we observe no mixing between the positive and negative helicity Weyl components $\Psi_L^{+,(1)}$ and $\Psi_L^{-,(2)*}$. Our current understanding is that $\Psi_L^{+,(1)}$ and its complex conjugate are the boundary data on $\scri$ and, consequently, $\Psi_L^{-,(2)}$ should not be present on the phase space. A simple counting of degrees of freedom for the left handed Weyl spinor supports our results.
    
    \item In sections \ref{Sec:ED} and \ref{Sec:YM}, we exhaustively explored asymptotic symmetries for gauge theories in decelerating spatially flat FLRW. In particular, we highlighted a severe inconsistency concerning the introduction of an ``extended phase space" at null infinity and the definition of the soft photon field required to connect asymptotic symmetries with soft theorems and memory effects. In order to construct an invertible symplectic form, we had to assume that the bulk and soft fields ($\hat{\cA}_B, N_B$ and $G_B$) are independent of each other. Nonetheless, the original definition of the soft photon as $N_B=\int_\R \p_u \cA_B \pd u=\int_\R\p_u \hat{\cA}_B\pd u$ indicates that such an assumption is, in fact, not rigorous.
    
    \item Within section \ref{Subsec:Memories}, we delved into the gauge memory effects in FLRW spacetimes. Our treatment shows that the soft photon field still generates the kick-memories in FLRW, although the momentum $p^A$ is replaced by its comoving version $P^A=a^2p^A$. 
    
    Furthermore, we explored the ordinary memory effect, by means of a derivation of Liénard-Wiechert-like solutions for spatially flat FLRW spacetimes (appendix \ref{LienardWiechertLikeSolutionsforFLRW}), as well as the null memory effect. We also noted that the displacement memory effect is altered with respect to flat spacetimes, such that the displacement cannot be directly related to the soft photon.
    
    Overall, we described in detail how the physical interpretation, regime of validity and setup of the different memories are modified with respect to flat spacetimes.

    \item In section \ref{subsec:fermionicps}, and making use of appendix \ref{AppendixExpansion}, we investigated the asymptotic behavior and phase space of a non-Abelian gauge field with dynamically coupled matter fields, i.e. conformally coupled complex scalar fields and Dirac fields, for an FLRW background. This is a challenging setting, both at a computational and at a conceptual level, which had not been explored before in the flat literature to the best of our knowledge. 
    
    Once the dust had settled, we attempted to construct the asymptotic phase space of the theory and realized about diffculties arising to conciliate the use of Poisson superbrackets and the covariant phase space formalism when dynamical fermions are present. Concretely, within the presented formalism, we can either disregard the phase space derivation via symplectic form of the charge algebra and adopt Poisson superbrackets directly or we can treat the phase space for bosons and fermions separately but the use of a single phase space seems inconsistent with the use of Poisson superbrackets and a closed asymptotic charge algebra.
    
\end{itemize}

\subsubsection*{Open questions and future guidelines}

Finally, we comment on open questions and future research directions:

\begin{itemize}
    \item First of all, we consider urgent the need to solve the aforementioned inconsistencies arising when trying to make the soft photon field definition compatible with the ``extended phase space" at null infinity. This is, in our understanding, merely a technical problem which is possibly caused by the fact that we are integrating over a null hypersurface (instead of a spacelike surface) when deriving the symplectic form and corresponding Poisson brackets. Nonetheless, taking into account that this issue has remained unnoticed in the literature so far and that it affects both flat and FLRW backgrounds, we firmly think that it should be properly addressed in future works.
    
    \item Extension to other FLRW universes such as closed and open, as well as to the more phenomenologically relevant inflationary ones, namely spatially flat and accelerating FLRW, is a logical step to pursue. Similarly to the gravitational analysis, a major obstacle which arises in the accelerating case is that the interesting null loci (particle and event horizons) are observer dependent.

    \item We encourage to deepen into further studies with the aim of unveiling new memory effects specific to cosmological backgrounds and to investigate the soft theorem corner, in order to complete the infrared triangle for gauge theories in FLRW.
    
    \item We find fascinating the recent connection between classical solutions of gravity and gauge theories, i.e. the classical double copy \cite{Monteiro:2014cda}, and the asymptotic structure of flat spacetimes \cite{Campiglia:2021srh,Adamo:2021dfg}. We hope that the results herein, together with the analysis of gravitational asymptotic symmetries in FLRW performed in \cite{Heckelbacher,Rojo,Olivieri}, can pave the way to an extension of the previous relationship to cosmological spacetimes.
    
    \item Finally, we find it especially interesting to continue aiming for a consistent covariant phase space formulation and symplectic form when fermions are present. While the approach presented in this paper did not lead to the desired result, an attempt would be a construction similar to the formalism used in \cite{Henneaux:1992ig}. Such a consistent formulation will also be required if we intend to transition from a purely classical analysis to a quantum one in the field.
\end{itemize}


\section*{Acknowledgements}
The authors would like to thank Ivo~Sachs for support, as well as Till~Heckelbacher for initial collaboration and stimulating discussions, Johannes~Pirsch for providing us with references regarding the inclusion of fermions on the phase space and Irina~Kharag for proofreading this paper. M.E.R. is grateful to Hr\'olfur \'Asmundsson for discussions on the classical double copy and its relevance in connecting gauge theories with gravity. T.S. thanks Orville Damaschke for insightful discussions on null-hypersurfaces. The work of M.E.R. was funded by the Excellence Cluster Origins of the DFG under Germany’s Excellence Strategy EXC-2094 390783311. The work of T.S. was partially supported by the DFG through the Research Training Group "GRK 2149: Strong and Weak Interactions - from Hadrons to Dark Matter".


\appendix

\section{Spinor conventions}
\label{Appendix:Spinorconventions}
For our discussion, we will choose the $\gamma$-matrices in the chiral (Weyl) basis
\begin{align}  
\gamma^{\Bar{\mu}}=\begin{pmatrix} 0 & \sigma^{\Bar{\mu}} \\ \bar{\sigma}^{\Bar{\mu}} & 0
\end{pmatrix} && \text{with} && \sigma^{\Bar{\mu}}=\left(\mathbb{I}_2, \boldsymbol{\sigma}\right), \;\;\; \bar{\sigma}^{\Bar{\mu}}=\left(\mathbb{I}_2,- \boldsymbol{\sigma}\right)
\end{align}
with the Pauli matrices $\boldsymbol{\sigma}=\left(\sigma^{\Bar{1}}, \sigma^{\Bar{2}}, \sigma^{\Bar{3}}\right)$
\begin{align}
    \sigma^{\Bar{1}}=\begin{pmatrix} 0 & 1\\ 1& 0\end{pmatrix}, && \sigma^{\Bar{2}}=\begin{pmatrix} 0 & -i \\ i & 0\end{pmatrix}, && \sigma^{\Bar{3}}=\begin{pmatrix} 1 & 0 \\ 0 & -1\end{pmatrix}.
\end{align}
Barred indices are used to denote Lorentz indices in distinction to spacetime indices.\\
Next, we consider an infinitesimal Lorentz transformation $\tensor{\Lambda}{^{\bar{\mu}}_{\bar{\nu}}}=\delta^{\bar{\mu}}_{\bar{\nu}}+\tensor{\lambda}{^{\bar{\mu}}_{\bar{\nu}}}$ which preserves the Minkowski metric
$\tensor{\Lambda}{^{\bar{\mu}}_{\bar{\nu}}}\tensor{\Lambda}{^{\bar{\rho}}_{\bar{\sigma}}}\eta_{\bar{\mu}\bar{\rho}}=\eta_{\bar{\nu}\bar{\sigma}}$ by definition. A Dirac spinor field transforms under an infinitesimal Lorentz transformation according to
\begin{align}
    \Psi(y)\longmapsto \Psi'( y)=\underbrace{\left(1+\frac{1}{4}\sigma^{{\bar{\mu}}\bar{\nu}} \lambda_{\bar{\mu}\bar{\nu}}\right)}_{\equiv S(\Lambda)}\Psi(\Lambda^{-1} y) \ , \label{spinortrafo}
\end{align}
where we omitted spinor indices and $\sigma^{\bar{\mu}\bar{\nu}}=\frac{1}{2}[\gamma^{\bar{\mu}}, \gamma^{\bar{\nu}}]$.
For the generalization to the FLRW spacetime $\mathcal{M}$ \eqref{Metric}, we introduce a tetrad $e^{\Bar{\alpha}}_\mu(x)$
such that the position dependence of the metric can be expressed in terms of the tetrad as
\begin{align}
    g_{\mu\nu}(x)=e^{\Bar{\alpha}}_\mu(x) e^{\Bar{\beta}}_\nu(x) \eta_{\Bar{\alpha}\Bar{\beta}} \ ,
\end{align}
with $\eta_{\bar{\alpha}\bar{\beta}}=\diag(1,-1,-1,-1)$. We choose the tetrad such that the only non-vanishing components are
\begin{align}
e^{\bar{0}}_u=a(u,r), && e^{\bar{0}}_r=a(u,r), && e^{\bar{1}}_r=a(u,r), && e^{\bar{2}}_{\theta}=a(u,r)r, && e^{\bar{3}}_{\varphi}=a(u,r)r\sin(\theta), \\
e^u_{\bar{0}}=\frac{1}{a(u,r)}, && e^u_{\bar{1}}=-\frac{1}{a(u,r)}, && e^{r}_{\Bar{1}}=\frac{1}{a(u,r)}, && e^\theta_{\Bar{2}}=\frac{1}{a(u,r)r}, && e^\varphi_{\Bar{3}}=\frac{1}{a(u,r)r\sin(\theta)}.
\end{align}
We introduce now a spin connection
\begin{align}
    \nabla_{\Bar{\alpha}}\Psi=e_{\Bar{\alpha}}^\mu \left[\p_\mu + \frac{1}{4}\sigma^{\Bar{\beta}\bar{\gamma}}\omega_{\Bar{\beta}\Bar{\gamma}\mu} \right]\Psi && \text{with} && \omega_{\Bar{\alpha}\Bar{\beta}\mu}=e_{\Bar{\alpha}}^\nu \left(\p_\mu e_{\Bar{\beta}\nu}-\Gamma_{\mu\nu}^\rho e_{\Bar{\beta}\rho}\right)
\end{align}
such that the spin connection components (up to those related via the symmetry $\omega_{\Bar{\alpha}\Bar{\beta}\mu}=-\omega_{\Bar{\beta}\Bar{\alpha}\mu}$) for $\mathcal{M}$ are
\begin{align}
    \omega_{\Bar{0}\Bar{1}r}=\frac{s}{u+r}, && \omega_{\Bar{0}\Bar{2}\theta}=\frac{sr}{u+r}, && \omega_{\Bar{0}\Bar{3}\varphi}=\frac{sr\sin(\theta)}{u+r}, \nonumber\\
    \omega_{\Bar{1}\Bar{2}\theta}=1, && \omega_{\Bar{1}\Bar{3}\varphi}=\sin(\theta), && \omega_{\Bar{2}\Bar{3}\varphi}=\cos(\theta). \label{spinconnection}
\end{align}

\section{Asymptotic expansion of Yang-Mills with dynamical matter Fields\label{AppendixExpansion}}

In this appendix, we consider the asymptotic expansion of the equations of motion arising from the action \eqref{fullaction}. We use the following expansions for the fields
\begin{footnotesize}
\begin{align}
    \bpsi_{L,R}^+(u,r,\bth)=\frac{1}{a^{\frac{3}{2}}(u,r)} \sum_{n=1}^\infty \frac{\bpsi_{L,R}^{+,(n)}(u,\bth)}{r^n}, && \bpsi_{L,R}^-(u,r,\bth)=\frac{1}{a^{\frac{3}{2}}(u,r)}\sum_{n=2}^\infty \frac{\bpsi_{L,R}^{-,(n)}(u,\bth)}{r^n},\\
    \bphi(u,r,\bth)=\frac{1}{a(u,r)}\sum_{n=1}^\infty \frac{\bphi^{(n)}(u,\bth)}{r^n}, &&   A_u(u,r,\bth)=\sum_{n=1}^{\infty} \frac{A^{(n)}_u(u,\bth)}{r^n}, \\ A_r(u,r,\bth)=0, && A_B(u,r,\bth)=\sum_{n=0}^\infty \frac{A_B^{(n)}(u,\bth)}{r^n}.
\end{align}
\end{footnotesize}
The definitions for the spinor components are the same as defined around equation \eqref{plusminusspinor}. For $n\geq 1$, the Yang-Mills equations read:
\begin{footnotesize}
\begin{align}
    -\p_u A_u^{(1)\fa}+\gamma^{AB}D_A \p_u A_B^{(0)\fa}+ic\gamma^{AB} \llbracket A_A^{(0)}, \p_u A_B^{(0)}\rrbracket^\fa-ic\left(\p_u \bphi^{(1)\dagger} T^\fa\bphi^{(1)}-\bphi^{(1)\dagger}T^\fa \p_u \bphi^{(1)}\right)+\nonumber\\ +c\left(\bpsi_L^{+,(1)\dagger}T^\fa \bpsi_L^{+,(1)}+\bpsi_R^{+,(1)\dagger}T^\fa \bpsi_R^{+,(1)}\right)=0, \label{965} \\
    -n(n-1)A_u^{(n)\fa}-D^2A_u^{(n)\fa}-(n+1)\p_u A_u^{(n+1)\fa}+\gamma^{AB}D_A\p_u A_B^{(n)\fa}+\nonumber\\+ic\sum_{\substack{k+m=n\\k\geq 0, m\geq 1}}\left\{\gamma^{AB}D_A\llbracket A_B^{(k)}, A_u^{(m)}\rrbracket^\fa+ m \llbracket A_u^{(k+1)}, A_u^{(m)}\rrbracket^\fa +\gamma^{AB}\llbracket A_A^{(k)}, \p_B A_u^{(m)}\rrbracket^\fa \right\}+\nonumber\\
    +ic\sum_{\substack{k+m=n\\k\geq 0, m\geq 0}}\Bigl\{ \gamma^{AB}\llbracket A_A^{(k)}, \p_u A_B^{(m)}\rrbracket^\fa-\p_u \bphi^{(k+1)\dagger} T^\fa\bphi^{(m+1)}+\bphi^{(k+1)\dagger}T^\fa \p_u \bphi^{(m+1)}-\nonumber \\-i\left(\bpsi_L^{+,(k+1)\dagger}T^\fa \bpsi_L^{+,(m+1)}+\bpsi_R^{+,(k+1)\dagger}T^\fa\bpsi_R^{+,(m+1)}\right)\Bigr\}
    + c^2\sum_{\substack{k+m+l=n\\k,l\geq 0; m\geq 1}}\Bigl\{ \gamma^{AB}\llbracket A_A^{(k)}, \llbracket A_B^{(l)}, A_u^{(m)}\rrbracket\rrbracket^\fa+
    \nonumber\\+\bphi^{(k+1)\dagger}\left(T^\fa A_u^{(m)}+A_u^{(m)}T^\fa\right)\bphi^{(l+1)}\Bigr\}-ic\sum_{\substack{k+m+1=n\\k\geq 0, m\geq 1}}i\left(\bpsi_L^{-,(k+2)\dagger}T^\fa \bpsi_L^{-,(m+1)}+\bpsi^{-,(k+2)\dagger}_RT^\fa\bpsi_R^{-,(m+1)}\right)=0,\label{966}\\
    -n(n+1)A_u^{(n+1)\fa}-n\gamma^{AB}D_A A_B^{(n)\fa}+ic\sum_{\substack{k+m=n\\k\geq 0, m\geq 1}} \Bigl\{m\gamma^{AB}\llbracket A_A^{(k)}, A_B^{(m)}\rrbracket^\fa-2i\Bigl(\bpsi_L^{-,(k+2)\dagger}T^\fa\bpsi_L^{-,(m+1)}+\nonumber\\+\bpsi_R^{-,(k+2)\dagger}T^\fa\bpsi_R^{-,(m+1)}\Bigr)\Bigr\}+ic\sum_{\substack{k+m=n\\k,m\geq 0}} (k-2m)\bphi^{(k+1)\dagger}T^\fa \bphi^{(m+1)}=0,\label{967}\\
    -2\p_u A_C^{(1)\fa}+\p_C A_u^{(1)\fa}+ic\llbracket A_u^{(1)}, A_C^{(0)}\rrbracket^\fa-D^2 A_C^{(0)\fa}+\gamma^{AB}D_A D_C A_B^{(0)\fa}+ic \gamma^{AB}D_A\llbracket A_B^{(0)}, A_C^{(0)}\rrbracket^\fa+\nonumber\\+ic\gamma^{AB}\llbracket A_A^{(0)}, \p_B A_C^{(0)}-\p_C A_B^{(0)}\rrbracket^\fa+c^2 \gamma^{AB}\llbracket A_A^{(0)}, \llbracket A_B^{(0)}, A_C^{(0)}\rrbracket\rrbracket^\fa-ic\bigl( \p_C \bphi^{(1)\dagger}T^\fa \bphi^{(1)}-\bphi^{(1)\dagger} T^\fa \p_C \bphi^{(1)}\bigr)+\nonumber\\+c^2 \bphi^{(1)\dagger}\left(T^\fa A_C^{(0)}+A_C^{(0)}T^\fa\right)\bphi^{(1)}-j_C^{\Psi(2)\fa}=0,\label{968}\\
    -2(n+1)\p_u A_C^{(n+1)\fa}-n(n+1)A_C^{(n)\fa}+(n+1)\p_C A_u^{(n+1)\fa}-D^2 A_C^{(n)\fa}+\gamma^{AB}D_A D_C A_B^{(n)\fa}+\nonumber\\+ic\sum_{\substack{k+m=n\\k,m\geq 0}}\Bigl\{(n+1)\llbracket A_u^{(k+1)}, A_C^{(m)}\rrbracket^\fa+\gamma^{AB}D_A \llbracket A_B^{(k)}, A_C^{(m)}\rrbracket^\fa+\gamma^{AB}\llbracket A_A^{(k)}, \p_B A_C^{(m)}-\p_C A_B^{(m)}\rrbracket^\fa-\nonumber\\-\left(\p_C \bphi^{(k+1)\dagger}T^\fa \bphi^{(m+1)}-\bphi^{(k+1)\dagger}T^\fa\p_C \bphi^{(m+1)}\right)+ic\sum_{\substack{k+m=n\\k\geq 1,m\geq 0}} m \llbracket A_u^{(k)}, A_C^{(m)}\rrbracket^\fa-j_C^{\Psi(n+2)\fa}+\nonumber\\+c^2\sum_{\substack{k+m+l=n\\k,m,l\geq 0}}\left\{\gamma^{AB}\llbracket A_A^{(k)}, \llbracket A_B^{(l)}, A_C^{(m)}\rrbracket\rrbracket^\fa+\bphi^{(k+1)\dagger}\left(T^\fa A_C^{(l)}+A_C^{(l)}T^\fa\right)\bphi^{(m+1)}\right\}=0 , \label{969}
    \end{align}
\end{footnotesize}
where $j_C^{\Psi\fa}=-c\bar{\bzeta}\frac{\gamma_C}{a}T^\fa \bzeta$ and hence for $n\geq 0$
\begin{footnotesize}
\begin{align}\label{97071}
    j_\theta^{\Psi(n+2)\fa}=ic\sum_{\substack{k+m=n\\ k,m\geq 0}} \Bigl(\bpsi_L^{+,(k+1)\dagger}T^\fa \bpsi_L^{-,(m+2)}-\bpsi_L^{-,(m+2)\dagger}T^\fa\bpsi_L^{+,(k+1)}+\nonumber\\+\bpsi_R^{+,(k+1)\dagger}T^\fa \bpsi_R^{-,(m+2)}-\bpsi_R^{-,(m+2)\dagger}T^\fa\bpsi_R^{+,(k+1)}\Bigr),\\
    j_\varphi^{\Psi(n+2)\fa}=c\sin(\theta)\sum_{\substack{k+m=n\\k,m\geq 0}}\Bigl(\bpsi_R^{+,(k+1)\dagger}T^\fa \bpsi_R^{-,(m+2)}+\bpsi_R^{-,(m+2)\dagger}T^\fa \bpsi_R^{+,(k+1)}-\nonumber\\-\bpsi_L^{+,(k+1)\dagger}T^\fa \bpsi_L^{-,(m+2)}-\bpsi_L^{-,(m+2)\dagger}T^\fa\bpsi_L^{+,(k+1)}\Bigr) .
\end{align}
\end{footnotesize}
For $n\geq 1$, the Klein-Gordon equation becomes
\begin{footnotesize}
\begin{align}\label{972}
    -2n\p_u \bphi^{(n+1)}-n(n-1)\bphi^{(n)}-D^2\bphi^{(n)}+ic\sum_{\substack{m+k=n\\k\geq 1, m\geq 0}} \Bigl\{2nA_u^{(m+1)}\bphi^{(k)}+\nonumber\\+\gamma^{AB}\left(2A_A^{(m)}D_B\bphi^{(k)}+D_A A_B^{(m)}\bphi^{(k)}\right)\Bigr\}+c^2\sum_{\substack{k+l+m=n\\k\geq 1; l,m\geq 0}}\gamma^{AB}A_A^{(l)}A_B^{(m)}\bphi^{(k)}=0.
\end{align}
\end{footnotesize}
For $n\geq 1$, the Dirac equation takes the form
\begin{footnotesize}
\begin{align}
    n\bpsi_L^{+,(n+1)}+\left(\frac{\p_\varphi}{\sin(\theta)}-i\p_\theta-\frac{i}{2\tan(\theta)}\right)\bpsi_L^{-,(n+1)}-ic\sum_{\substack{k+m=n\\k\geq 1, m\geq 0}}\left(\frac{A_\varphi^{(m)}}{\sin(\theta)}-iA_\theta^{(m)}\right)\bpsi_L^{-,(k+1)}=0,\label{973}\\
    2\p_u \bpsi_L^{-,(n+1)}+(n-1)\bpsi_L^{-,(n)}-2ic\sum_{\substack{k+m+1=n\\k\geq 1, m\geq 0}}A_u^{(m+1)}\bpsi_L^{-,(k+1)}-\nonumber\\-\left(\frac{\p_\varphi}{\sin(\theta)}+i\p_\theta+\frac{i}{2\tan(\theta)}\right)\bpsi_L^{+,(n)}+ic\sum_{\substack{k+m=n\\k\geq 1, m\geq 0}} \left(\frac{A_\varphi^{(m)}}{\sin(\theta)}+iA_\theta^{(m)}\right)\bpsi_L^{+,(k)}=0,\label{974}\\
    n\bpsi_R^{+,(n+1)}-\left(\frac{\p_\varphi}{\sin(\theta)}+i\p_\theta+\frac{i}{2\tan(\theta)}\right)\bpsi_R^{-,(n+1)}+ic\sum_{\substack{k+m=n\\k\geq 1, m\geq 0}}\left(\frac{A_\varphi^{(m)}}{\sin(\theta)}+iA_\theta^{(m)}\right)\bpsi_R^{-,(k+1)}=0,\label{975}\\
    2\p_u \bpsi_R^{-,(n+1)}+(n-1)\bpsi_R^{-,(n)}-2ic\sum_{\substack{k+m+1=n\\k\geq 1, m\geq 0}}A_u^{(m+1)}\bpsi_R^{-,(k+1)}+\nonumber\\+\left(\frac{\p_\varphi}{\sin(\theta)}-i\p_\theta-\frac{i}{2\tan(\theta)}\right)\bpsi_R^{+,(n)}-ic\sum_{\substack{k+m=n\\k\geq 1, m\geq 0}}\left(\frac{A_\varphi^{(m)}}{\sin(\theta)}-iA_\theta^{(m)}\right)\bpsi_R^{+,(k)}=0.\label{976}
\end{align}
\end{footnotesize}
Now, we want to determine the data that needs to be specified in order to find unique solutions to the above equations to all orders. Let us begin with $\cA_B, \bphi^{(1)}, \bpsi_{L,R}^{+,(1)}$ \footnote{We need also the Hermitian conjugates of the matter fields which we do not explicitly write for brevity.}. Then, \eqref{965} determines $A_u^{(1)}$ uniquely by specifiying it at a certain value of $u$. With this, \eqref{974} determines $\bpsi_L^{-,(2)}$ and \eqref{976} determines $\bpsi_R^{-,(2)}$ uniquely upon fixing their values for one specific $u$. Thereby, \eqref{97071} yields $j_C^{\Psi(2)}$ which allows via \eqref{968} to determine $A_C^{(1)}$ upon specifying its value for a specific $u$. Using \eqref{972}, we can determine now $\bphi^{(2)}$ completely, by fixing its value for a specific $u$. Moreover, \eqref{973} and \eqref{975} unambigiously determine $\bpsi_L^{+,(2)}$ and $\bpsi_R^{+,(2)}$, respectively.  Afterwards, we can move on as follows: \eqref{967} fully fixes $A_u^{(2)}$, \eqref{974} and \eqref{976} respectively yield $\bpsi_L^{-,(3)}$ and $\bpsi_R^{-,(3)}$ if we specify the latter two at a certain $u$. From \eqref{97071}, we can then obtain $j_C^{\Psi(3)}$ which in turn allows via \eqref{969} to determine $A_C^{(2)}$ upon specifying its value at a certain $u$. With \eqref{972}, we can then determine $\bphi^{(3)}$ by fixing its value at a certain $u$. Moreover, \eqref{973} and \eqref{975} give us then 
$\bpsi_L^{+,(3)}$ and $\bpsi_R^{+,(3)}$.\\
Thereafter, we can run through a loop, beginning
from \eqref{967} and moving on in the order \eqref{974}, \eqref{976}, \eqref{97071}, \eqref{969}, \eqref{972}, \eqref{973}, \eqref{975} and closing the loop by starting again from \eqref{967}. Thus, given $\cA_B, \bphi^{(1)}, \bpsi_{L,R}^{+,(1)}$, we can uniquely determine the fields to any order in $r$, by specifying their values at a certain value of $u$ where necessary (also for any order). These constitute the boundary data of our theory.

\section{Liénard-Wiechert-like solutions for FLRW}
\label{LienardWiechertLikeSolutionsforFLRW}
In this appendix, we find an expression for the electromagnetic field strength generated by a massive charged point particle for FLRW backgrounds. The corresponding solutions in flat spacetime are known as Liénard-Wiechert solutions. We closely follow the derivation in flat spacetime \cite{Jackson}. \\
The Maxwell equations for a FLRW background take the form
\begin{align}
    g^{\lambda\mu}_{\mathbb{M}}\widehat{\nabla}_\lambda F_{\mu\nu}=a^2 J_\nu.
\end{align}
We use now the non-covariant flat Lorenz gauge condition $\p_\eta A_\eta -\p_x A_x -\p_y A_y -\p_z A_z=0$ and obtain 
\begin{align}
    (\p_\eta^2- \p_x^2-\p_y^2-\p_z^2)A_{\mu}=a^2 J_{\mu}.
\end{align}
Up to the scale factor and use of conformal time, this is just the same equation we would find when looking at a Minkowski background. The Green's function
\begin{align}
    \mathcal{G}(x)=\frac{\vartheta(\eta)}{4\pi r}\delta(r-\eta),
\end{align}
where $\vartheta$ denotes the Heaviside step function, satisfies $(\p_\eta^2- \p_x^2-\p_y^2-\p_z^2)\mathcal{G}(x)=\delta^{(4)}(x)\equiv \delta(\eta)\delta^{(3)}(\bx)$. The wave equation is correspondingly solved by
\begin{align}
    A_\mu(x)&= \int_0^\infty \int_{\R^3} \frac{\vartheta(\eta-\eta')}{4\pi|\bx-\bx'|} \delta(|\bx-\bx'|-(\eta-\eta')) a^2(\eta') J_\mu(x') \pd \eta' \pd^3 x'=\nonumber\\&=
\int_0^\infty \int_{\R^3} \frac{\vartheta(\eta-\eta')}{2\pi} \delta\left((\eta-\eta')^2-(\bx-\bx')^2\right) a^2(\eta') J_\mu(x') \pd \eta' \pd^3 x'.
\end{align}
We consider a massive point particle of electric charge $Q$ as the source of the charge current. The particles' four-velocity is $u^\mu(\tau)=\frac{\pd x_Q^\mu(\tau)}{\pd \tau}=(u^\eta, \boldsymbol{u})$ where $x^\mu_Q(\tau)=(\eta_Q(\tau), \bx_Q(\tau))$ and $\tau$ denotes the proper time along its worldline. The charge current is given by
\begin{align}
    J_\mu(x)=Q\int  u_\mu(\tau) \frac{1}{a^4(\eta)} \delta^{(4)}(x-x_Q(\tau)) \pd\tau
\end{align}
and, therefore, the gauge field becomes
\begin{align}
    A_\mu(x)=\frac{Q}{2\pi}\int \vartheta(\eta-\eta_Q) \delta\left((\eta-\eta_Q)^2-R^2\right) \frac{u_\mu(\tau)}{a^2(\eta_Q)} \pd\tau \ ,
\end{align}
where we denoted $\bR=\bx-\bx_Q$ and $R=|\bR|$ to simplify the equation.\\
Henceforth, we will assume $R\neq 0$. This makes sense because we want to apply the solutions to cases for which the radiation source, i.e. the charge $Q$, is far away from the observer. In this case, the derivative of the $\vartheta$-function does not contribute, since it would lead to an evaluation of the integrand at $R=0$, and one finds
\begin{align}
    \p_\mu A_\nu(x)=\frac{Q}{2\pi}\int  \vartheta(\eta-\eta_Q) \p_\mu \delta\left((\eta-\eta_Q)^2-R^2\right)\frac{u_\mu(\tau)}{a^2(\eta_Q)}\pd \tau.
\end{align}
Next, we define for convenience $f(\tau)=g_{\mathbb{M},\alpha\beta}(x-x_Q)^\alpha (x-x_Q)^\beta$ which is just the argument of the above Dirac-$\delta$. Generally, we have $\p_\mu \delta(f(\tau))=\p_\mu f \frac{\pd \tau}{\pd f} \frac{\pd}{\pd \tau}\delta(f(\tau))$. Explicitly, we find
\begin{align}
    \p_\mu \delta(f(\tau))=-\frac{g_{\mathbb{M},{\mu\gamma}}(x-x_Q)^\gamma}{g_{\mathbb{M},\alpha\beta}(x-x_Q)^\alpha u^\beta} \frac{\pd}{\pd \tau}\delta(f(\tau)).
\end{align}
Plugging this into the previous expression, we obtain via integration by parts
\begin{align}
    \p_\mu A_\nu(x)&= \frac{Q}{2\pi} \int \vartheta(\eta-\eta_Q) \frac{\pd}{\pd \tau}\left(\frac{u_\nu(\tau)}{a^2(\eta_Q)}\frac{g_{\mathbb{M},{\mu\gamma}}(x-x_Q)^\gamma}{g_{\mathbb{M},\alpha\beta}(x-x_Q)^\alpha u^\beta}\right) \delta(f(\tau)) \pd \tau.
\end{align}
Finally we have to rewrite the Dirac-$\delta$ as $\delta(f(\tau))=\frac{\delta(\tau-\tau_\text{ret})}{2g_{\mathbb{M},\alpha\beta}(x-x_Q)^\alpha u^\beta}$ where $\tau_\text{ret}$ is defined by $f(\tau_\text{ret})=0$. The physical meaning of this condition is that the observer only sees radiation from the source if the events of emission and observation are light-like separated. Equivalently, we can write
\begin{align}
    \eta_Q(\tau_\text{ret})=\eta-|\bx-\bx_Q(\tau_\text{ret})|.
\end{align}
Hence, one obtains with the previous expression for the field strength
\begin{align}
F_{\mu\nu}(x)=\frac{Q}{4\pi} \frac{1}{g_{\mathbb{M},\lambda\kappa}(x-x_Q)^\lambda u^\kappa}  \frac{\pd}{\pd \tau}\left( \frac{(x-x_Q)^\gamma\left(  u_\nu(\tau) g_{\mathbb{M},{\mu\gamma}}-u_\mu(\tau) g_{\mathbb{M},{\nu\gamma}}\right)}{a^2(\eta_Q) g_{\mathbb{M},\alpha\beta}(x-x_Q)^\alpha u^\beta}\right)\bigg\vert_{\tau=\tau_\text{ret}}.
\end{align}
For $a(\eta)=1$, this becomes the Liénard-Wiechert solution \cite{Jackson}, such that we consistently recover the correct flat limit. \\
Since the indices could lead to the erroneous impression that the above expression was covariantly derived, we introduce a notation which captures the non-covariance more appropriately. We write for the (rescaled) peculiar velocity $\scriptstyle\bv_Q\equiv a(\eta_Q)\frac{\pd \bx_Q}{\pd t_Q}$ and $\scriptstyle \gamma\equiv \frac{1}{\sqrt{1-\bv_Q^2}}$. Accordingly, we have $ \scriptstyle
    g_{\mathbb{M},\lambda\kappa}(x-x_Q)^\lambda u^\kappa=\frac{\gamma(\eta_Q)}{a(\eta_Q)}\left(\eta-\eta_Q-\bR\bv_Q\right).$
Furthermore, we are only interested in the $\eta i$-components of the field strength:
{\small
\begin{align}
    F_{\eta i}(x)=\frac{Q}{4\pi(R-\bR\bv_Q)^2}\biggl\{ \frac{1}{\gamma^2} \frac{\bR^i-R\bv_Q^i}{R-\bR\bv_Q}+a(\eta_Q)\left(\frac{\bR\dot{\bv}_Q}{R-\bR\bv_Q}(\bR^i-R\bv_Q^i)-R\dot{\bv}_Q^i\right)\biggr\}\bigg\vert_{\eta_Q=\eta-R(\eta_Q)} \label{FRI},
\end{align}}
where $\dot{f}=\frac{\pd}{\pd t}f$.

\bibliography{references}
\bibliographystyle{JHEP}

\end{document}